\documentclass[prc,aps,floatfix,onecolumn,groupedaddress,nofootinbib,showpacs,preprintnumbers,superscriptaddress,
amsmath,amssymb,amsfonts,widetable]{revtex4}
\usepackage{array}
\usepackage{hyperref}
\usepackage[nameinlink,capitalize]{cleveref}
\crefname{section}{Sec.}{Secs.}
\crefname{equation}{Eq.}{Eqs.}
\crefname{eqnarray}{Eq.}{Eqs.}
\crefname{table}{Table}{Tables}
\usepackage{xcolor}
\usepackage{graphicx}
\usepackage{bm}
\usepackage{soul}
\usepackage{array}
\usepackage{lipsum}
\usepackage{relsize}
\usepackage{mathrsfs}
\usepackage{amssymb}
\usepackage{amsmath}
\usepackage{graphicx}
\usepackage{booktabs}


\begin{document}

\title{Spin-2 Mesons in a Relativistic Hartree Description of Nuclei}
\author{Brendan T. Reed}
\affiliation{
Theoretical Division, Los Alamos National Laboratory, Los Alamos, NM 87545, USA
}
\author{Marc Salinas}
\affiliation{Lawrence Livermore National Laboratory, Livermore, CA 94550}
\preprint{LA-UR-25-30960}
\date{\today}

\begin{abstract}
    The role of the isovector spin-orbit potential in nuclear modeling has recently been explored in greater detail due largely in part to the PREX and CREX parity violating electron scattering (PVES) experiments. 
    The result of both experiments have shown that the neutron-rich skins of $^{48}$Ca and $^{208}$Pb, as determined by the experimental analysis, are largely incompatible with current nuclear model predictions. 
    One way to address, and potentially solve, this incompatibility has been to enhance the isovector spin-orbit sector of density functional models.
    Here, we explore the range of possible enhancements in the context of covariant density functional theory by introducing a new class of spin-2 \textit{massive} mesons to the Lagrangian. 
    In doing so, we find that these mesons not only meaningfully contribute to the theory, but also mitigate problems seen in other works which enhance spin-orbit effects in finite nuclei. 
    We also discuss the implications and future regarding this ``dilemma'' as it pertains to studies of nuclear forces and future experiments such as the Mainz Radius Experiment (MREX).
\end{abstract}
\maketitle

\section{Introduction}

The nuclear-physics community has taken a multi-faceted approach over the last few decades in trying to understand the structure of the atomic nucleus. 
Early experiments done by Rutherford and Chadwick provided the nucleonic picture of the nucleus, consisting of neutrons and protons \cite{Rutherford,Chadwick:1932}. 
At the same time, theoretical efforts were undertaken to describe basic nuclear-structure properties such as the binding energy of the nucleus. 
This gave rise to some of the first phenomenological models like the well-known liquid-drop (Bethe-Weiszacker) formula and nuclear shell model, the latter of which is still used today. 
Further experimental progress was made with elastic electron scattering of various nuclei at the Stanford Linear Accelerator (SLAC) carried out by Hofstadter \cite{Hofstadter:1956qs,Ehrenberg59}.
This led to using the first Born approximation to estimate the nuclear RMS charge radius and even the experimental charge density for $^{12}$C.
On the theoretical front, Johnson and Teller had just employed the first use of classical field theory in 1955 to describe the atomic nucleus, with Duerr then implementing the same theory covariantly a year after \cite{Johnson:1955zz,Duerr:1956zz}, thus, laying the groundwork for relativistic mean-field (RMF) theory \cite{walecka:1974} which has become a popular density functional theory (DFT) ever since, ranging in applicability from nuclear structure \cite{Fattoyev:2010mx,Dutra:2014,Reed_2021p2} to astrophysics \cite{Shen:2011kr,big-apple:2020}. 

In the present day, new experiments paralleling the work of Hofstadter have been conducted at Jefferson National Laboratory, namely PREX \cite{PREX,PREX2} and CREX \cite{CREX} for the Lead Radius Experiment and Calcium Radius Experiment, respectively. 
These campaigns provided novel information regarding the distribution of neutrons in the atomic nucleus, giving rise to the neutron-rich skin thickness that has been of continued debate as of recent.
The neutron-skin thickness, defined as the difference between the RMS radius of the neutron and proton ($R_\textrm{n} - R_\textrm{p})$, was inferred to be $R_\textrm{skin}^{208} = 0.283 \pm 0.071 \ \textrm{fm}$ for $^{208}$Pb and $R_\textrm{skin}^{48} = 0.121 \pm 0.035 \ \textrm{fm}$ for $^{48}$Ca \cite{PREX2,CREX}.
This caused a stir in the nuclear-physics community, since no consistent theoretical framework at the time could reconcile these two measurements at the $68\%$ confidence level due to the strong correlation between the neutron skin and the slope of the symmetry energy ($L_0$) \cite{Horowitz:2000xj}. 
Recent efforts from many-body approaches, for example utilizing Chiral Effective Field-Theory ($\chi$EFT), have offered strong ab-initio predictions for the slope of the symmetry energy \cite{Tews:2012fj,Tews:2018chv,Heinz2020_MSc,Hoppe2022PRC_IMSRG_IT,Alp:2025,Semposki:2025}, suggesting the nuclear equation of state is soft, providing more stringent constraints to mean field approaches.
Additionally, many state-of-the-art $\chi$EFT calculations of neutron-rich nuclei also show disagreement with the combined PREX+CREX results, despite predicting results consistent with other experimental nuclear structure observables \cite{Arthuis2024arxiv_LowResForces,Miyagi:2026}.

Recently, much progress has been made in exploring more exotic density functionals with the addition of the scalar-isovector $\delta$-meson \cite{Reed:2023}, tensor couplings \cite{salinas:2024,Kunjipurayil:2025xss,reed:2025}, and enhanced spin-orbit couplings \cite{Reed:2023,salinas:2024,Yue:2024}.
These are done with the purpose of modifying the isovector sector of the DFT in hopes to better explain the CREX+PREX results without sacrificing predictions of other nuclear structure properties. 
Recent evidence has suggested that the isovector spin-orbit interaction is responsible for the CREX/PREX discrepancy with non-relativistic models showing promising success in mitigating this dilemma \cite{Yue:2024,Kunjipurayil:2025xss}.
However, enhanced spin-orbit interactions within RMF theory are difficult to implement without affecting the other areas of the theory.
For example, as shown by \cite{Kunjipurayil:2025xss}, within the nonlinear RMF framework enhancing the spin-orbit sector disrupts the well-ordered structure of nuclear shells in neutron-rich doubly magic nuclei.
This is despite the model being shown to greatly improve the CREX+PREX ``dilemma'' by having a large neutron skin in $^{208}$Pb and a smaller skin in $^{48}$Ca.

Being a covariant field theory, the individual interactions between nucleons are governed by meson exchanges with the possibility of additional self- and cross-interactions among the mesons.
This presents a great difficulty in treating different components of the interaction separately due to the covariance between each piece of the interaction.
This is opposed to, for example, the hierarchal improvements to $\chi$EFT Hamiltonians which introduce additional interactions through increasing the number of terms in the low-energy chiral expansion \cite{Epelbaum2009RMP_ChiralEFTReview,Machleidt_2016}.
In contrast to $\chi$EFT, RMF theory lacks a systematic power-counting scheme that allows controlled, order-by-order improvements, as the hierarchy of mesonic degrees of freedom is model-dependent and constrained by the phenomenology of finite nuclei.
As a result, enhancing any one sector through the introduction of additional couplings without introducing new particles is difficult to implement.

However, we may introduce additional terms to the Lagrangian in a hierarchal way by introducing new mesons which interact with the desired degree of freedom without violating any of the previous conditions.
In this paper, we introduce a new class of mesons which are both an independent interaction from the other mesons already present \textit{and} meaningfully contribute to the spin-orbit sector of the RMF Lagrangian.
By incorporating a tensor structure into the new meson class, we also connect these particles to non-relativistic tensor forces as seen in other formalisms.
This paper will be structured as follows.
In \cref{sec:rmf} we introduce RMF theory in more detail and show how the individual particle interactions reduce to more familiar nuclear forces seen in non-relativistic theories such as $\chi$EFT.
We then introduce the new class of mesons and derive all equations of motion in the Hartree limit for finite nuclei in \cref{sec:tensor_meson_eqns}.
The impact and effects that these mesons have on finite nuclei are explored in \cref{sec:results}.
Finally, we discuss the implications of these meson interactions in future RMF studies and the possible alleviation of the PREX/CREX ``dilemma'' in \cref{sec:discussion} before concluding in \cref{sec:conclusions}.

\section{Relativistic Mean-Field Theory}
\label{sec:rmf}
\subsection{Relativistic Equations and the Mean-Field Approximation}
We begin first with a short introduction to the class of nonlinear relativistic models \cite{Lalazissis:1999,FSUGold,FSUGold2,Reed:2023,salinas:2024}.
The interaction Lagrangian for this class of models is
\begin{equation}
    \label{eq:lagrangian}
    \mathcal{L}_{\mathrm{int}}^0=\bar{\psi}\Big[\mathcal{S}(\phi,\bm{\delta})-\mathcal{V}_\mu(V_\mu,\bm{b_\mu},A_\mu)\gamma^\mu\Big]\psi-U(\phi)+\frac{\zeta}{4!}g_v^4(V_\mu V^\mu)^2+\Lambda_vg_v^2g_\rho^2V_\mu V^\mu\bm{b_\mu}\cdot\bm{b^\mu}
\end{equation}
where we write
\begin{eqnarray}\nonumber
    &&\mathcal{S}(\phi,\bm{\delta}) = g_s\phi+\frac{g_\delta}{2}\bm{\tau}\cdot\bm{\delta}\\
    &&\mathcal{V}_\mu(V_\mu,\bm{b_\mu},A_\mu) = g_vV_\mu+\frac{g_\rho}{2}\bm{\tau}\cdot \bm{b_\mu}+\frac{e}{2}(1+\tau_3)A_\mu\\\nonumber
    &&U(\phi) = \frac{\kappa}{3!}(g_s\phi)^3+\frac{\lambda}{4!}(g_s\phi)^4.
\end{eqnarray}
This class of models includes the interactions of nucleons ($\psi$) via the exchange of the isoscalar-scalar $\sigma$ ($\phi$), isoscalar-vector $\omega$ ($V_\mu$), isovector-vector $\rho$ ($\textbf{b}^\mu$), and isovector-scalar $\delta$ ($\bm{\delta}$) mesons.
In addition to interactions among the nucleons, this Lagrangian class also includes many self- and cross-coupling terms.
These are $\kappa$ and $\lambda$ for the cubic and quartic scalar $\phi$ self-coupling, $\zeta$ for the quartic $\omega$ self-coupling, and $\Lambda_v$ as the $\omega$-$\rho$ cross coupling.
The scalar self-coupling terms originally were added in to mimic the behavior of the three-nucleon interaction, an important effect in reducing the nuclear incompressibility ($K_{\rm sat}$) as well as reproducing the saturation properties of symmetric nuclear matter \cite{Glendenning:2000}.
The vector self-coupling is important for softening the high density behavior of nuclear matter, which is particularly important for reproducing several properties of neutron stars \cite{Fattoyev:2010rx}.
Finally the $\omega-\rho$ cross-coupling was introduced to help control the density dependence of the symmetry energy \cite{Horowitz:2000xj}.

In recent years, several model extensions have been considered and studied.
Largely motivated by the explosion of isovector-dependent measurements and the seemingly incompatible neutron skin measurements from the PREX \cite{PREX,PREX2} and CREX\cite{CREX} experiments, the $\delta$ meson was found to be important in softening the symmetry energy at low densities and was a meaningful addition to the theory \cite{Reed:2023}. 
Furthermore, tensor couplings were shown to improve the models as large Yukawa couplings for the $\delta$-meson created unphysical particle densities for finite nuclei \cite{salinas:2024} as well as expanding the spin-orbit force which is important in adjusting the charge radii of atomic nuclei \cite{Horowitz:2012we,Kunjipurayil:2025xss}.
These tensor couplings add the following to the Lagrangian of \cref{eq:lagrangian}
\begin{equation}
    \mathcal{L}_{\rm int}^{\rm T}=\bar{\psi} \sigma^{\mu\nu}\Big( \frac{f_v}{2M}\partial_\nu V_\mu+\frac{f_\rho}{2M}\bm{\tau}\cdot \partial_\nu\bm{B}_\mu\Big)\psi
    \label{eq:tensor_couplings}
\end{equation}
where $\sigma^{\mu\nu}=\frac{i}{2}[\gamma^\mu,\gamma^\nu]$ is the usual commutator of the Dirac matrices.

Taking the mean-field approximation, one may then write the previous covariant fields in terms of their classical expectation values.
The meson fields then reduce to classical Klein-Gordon fields with ground state proton and neutron densities as their source terms, where the single particle orbitals are solved from a Dirac Hamiltonian containing the interactions shown previously,
\begin{equation}
    \hat{H}=\bm{\alpha}\cdot \bm{p}+\beta\Big(M-S(r)\Big)+V(r)+i\bm{\gamma}\cdot\hat{r}\,T(r).
    \label{eq:hamiltonian}
\end{equation}
Here, the classical fields ($S(r),\, V(r)$) take the place of the covariant fields ($\mathcal{S}(\phi,\delta),\,\mathcal{V}_\mu(V_\mu,\bm{b_\mu},A_\mu)$) and the presence of the tensor couplings in \cref{eq:tensor_couplings} give rise to the field $T(r)$, defined below
\begin{subequations}
\begin{align}
S(r)  = &\Phi_{0}(r) + \frac{\tau_3}{2} \Delta_{0}(r), \\
V(r) = &W_{0}(r) + \frac{\tau_3}{2} B_{0}(r) + eA_{0}(r)\frac{(1\!+\!\tau_{3})}{2},\\
T(r) = &\frac{1}{2M}\left[\frac{f_{\rm v}}{g_{\rm v}}\frac{dW_{0}(r)}{dr} + \frac{\tau_3}{2} \frac{f_{\rho}}{g_{\rho}}\frac{dB_{0}(r)}{dr}\right].
\end{align}
\label{eq:mesons}
\end{subequations}
where for simplicity, we adopt the shorthand notation for the meson fields which combines the bare field with its respective Yukawa coupling,
\begin{center}
$(g_s\phi_0,g_vV_0,g_\rho b_0,g_\delta\delta_0)$
    $\rightarrow$
$(\Phi_0,W_0,B_0,\Delta_0)$.
\end{center}
Below, we now show how this relativistic Hamiltonian reduces in the non-relativistic limit, giving rise to familiar nuclear forces.

\subsection{Non-relativistic Reduction and Nuclear Forces}

To make the connection to specific nuclear forces, we may rewrite the relativistic equations by performing a non-relativistic reduction of the previous equations in the mean-field approximation following the work of \cite{Kunjipurayil:2025xss,Yue:2024}.
In doing so, one may make the substitution for the covariant Dirac bilinear $\psi$ to an auxiliary wave function $\mathcal{U}_{n\kappa}$.
To do so, we let the RMF Hamiltonian in \cref{eq:hamiltonian} act on $\psi$ to derive the Dirac equation.
The form of $\psi$ we take as
\begin{equation}
    \psi(x) = \frac{1}{r} \begin{pmatrix}
        g_{n \kappa}(r)\mathcal{Y}_{\kappa m}(\mathbf{\hat{r}})\\
        if_{n \kappa}(r)\mathcal{Y}_{-\kappa m}(\mathbf{\hat{r}})
    \end{pmatrix}.
    \label{eq:psi}
\end{equation}
where the upper and lower components are the wavefunctions $g_{n\kappa}$ and $f_{n\kappa}$, respectively.
One may obtain the auxiliary wave function $\mathcal{U}_{n\kappa}$ through the substitution
\begin{equation}
    \mathcal{U}_{n\kappa}(r)=\frac{g_{n\kappa}(r)}{\sqrt{\xi(r)}} \mathrm{, with} \,\,\,  \xi(r)=\Bigg(\frac{M^*(r)+E^*(r)}{M+E}\Bigg),
\end{equation}
where $M^*(r)=M-S(r)$ and $E^*(r)=E-V(r)$.
To make further progress, one must uncouple the upper and lower Dirac equations.
This can be done by rewriting the lower component in terms of the upper component, i.e.
\begin{equation}
    f_{n\kappa}(r) = \frac{1}{M^*(r)+E^*(r)}\Big(\frac{d}{dr}+\frac{\kappa^*(r)}{r}\Big)g_{n\kappa}(r).
\end{equation}
where $\kappa^*=\kappa+rT(r)$
Substituting this back into the Dirac equation for the upper component results in a second order differential equation for $g_{n\kappa}(r)$.
Following the work of \cite{darwin_term}, substituting $\mathcal{U}_{n\kappa}(r)$ into this equation eliminates all first derivatives resulting in an effective one-dimensional problem,
\begin{equation}
    \Big[\frac{d^2}{dr^2}-p^2-\frac{l(l+1)}{r^2}-U_{\rm eff}(r)\Big]\mathcal{U}_{n\kappa}(r)=0.
\end{equation}
For more details, we refer to \cite{Kunjipurayil:2025xss,darwin_term}.

The effective potential,
\begin{equation}
    U_{\rm eff}(r) = U_c(r)+(1+\kappa)U_{\rm so}(r)+U_D(r)+U_T(r)
\end{equation}
where subscripts $c$, so, $D$, and $T$ refer to the central, spin-orbit, Darwin \cite{darwin_term}, and tensor terms, respectively.
It is important to note that the tensor term only arises due to the presence of the tensor interaction in \cref{eq:tensor_couplings} and so it is not present in the regular FSUGold class of RMF models.
Here we write the form of the spin-orbit potential,
\begin{equation}
    U_{\rm so}(r) = \frac{1}{r(M^*(r)+E^*(r))}\frac{d}{dr}\Big(S(r)+V(r)\Big)-\frac{2}{r}T(r)
\end{equation}
with the meson potentials given in \cref{eq:mesons}.
Note that here there is an energy dependence for the nucleon potential, suppressed here by taking the non-relativistic limit where $E \approx M$.
In this way we can characterize an effective non-relativistic spin-orbit potential $V_\textrm{SO}(r) = \frac{U(r;E=M)}{2M}$.
We note that the potentials $S(r),V(r)$ and $T(r)$ appearing in the non-relativistic formulation are calculated using the ground-state Hartree solution of the nucleus.

We highlight here that there are two unique terms in this potential which contribute to the spin-orbit, a term containing the tensor interaction in \cref{eq:tensor_couplings} and another term which contains the mixed contribution of all scalar and vector mesons found in \cref{eq:lagrangian}.
The presence of this tensor term is the focus of some of this work since the tensor potential has a direct impact on the spin-orbit interaction. 
This way, one may utilize these tensor couplings to carefully tune the properties of the spin-orbit potential in nuclei.
Previous work including tensor interactions in relativistic mean field theory were largely done via the introduction of derivative couplings to the $\omega$ and $\rho$-mesons using \cref{eq:tensor_couplings}.
Some of these interactions were shown to slightly improve neutron-skin correlations while reducing the effects of large isovector meson fields, but largely limited the flexibility of the model since they are inherently tied to the $\omega$ and $\rho$-mesons and therefore not truly independent parameters \cite{salinas:2024,Kunjipurayil:2025xss, rufa:1988,Typel:2020ozc}.
It is the goal of this work to introduce new tensor-like interactions in RMF theory which are ideally explicitly independent of the other mesons.
In the next section, we introduce a new class of massive mesons which meaningfully contribute to the spin-orbit sector of nuclei and are explicitly independent of the other four mesons found within the FSUGold-class of RMF interactions.

\section{Massive Spin-2 Mesons}
\label{sec:tensor_meson_eqns}
To investigate both isoscalar and isovector tensor behaviors as seen in other nuclear theories, we consider the addition of tensor (spin-2) mesons to the traditional RMF Lagrangian of FSUGold-like models.
The fields representing such particles have a rank-2 tensor structure which we explore below.
For these particles, we couple them to the nucleonic fields ($\psi$) through the following Yukawa-like Lagrangian
\begin{equation}
    \mathcal{L}_{\rm int} = \bar{\psi} \sigma^{\mu\nu}\Big[g_TT_{\mu\nu} + \boldsymbol{\tau} \cdot g_H\mathbf{H}_{\mu\nu}\Big] \psi
\end{equation}
where $g_T$ and $g_H$ are the Yukawa couplings of the isoscalar tensor meson $T_{\mu \nu}$ and the isovector tensor meson $\mathbf{H}_{\mu \nu}$, respectively. 
We use the commutator $\sigma^{\mu \nu} = \frac{i}{2}[\gamma^\mu,\gamma^\nu]$ as in \cref{eq:tensor_couplings} to directly couple the tensor mesons to the nucleon fields.
Due to the antisymmetric and traceless features of $\sigma^{\mu\nu}$, the tensor fields $T_{\mu\nu}$ and $\mathbf{H}_{\mu \nu}$ must also be antisymmetric and traceless in order to contribute to the Lagrangian.

The kinetic energy contribution to the Lagrangian from these particles must follow a similar structure.
We write a general Lagrangian \cite{Suzuki1993,Ye2012,Dalmazi:2013cna,Koenigstein:2015asa} for spin-2 particles with these symmetries as
\begin{equation}
\begin{split}
    \mathcal{L}_{\rm kin} = \Big(-\frac{1}{2} m_T^2 T_{\mu\nu}T^{\mu\nu} +\frac{1}{8}T_{\mu\nu\sigma} T^{\mu\nu\sigma}\Big) +\Big(- \frac{1}{2} m_H^2 \mathbf{H}_{\mu\nu} \cdot \mathbf{H}^{\mu\nu} + \frac{1}{8} \mathbf{H}_{\mu\nu\sigma} \cdot \mathbf{H}^{\mu\nu\sigma}\Big)
\end{split}
\end{equation}
where we have defined the field strength tensors as,
\begin{equation}
\begin{split}
T_{\mu\nu\sigma} &= \partial_\mu T_{\nu \sigma}+\partial_\nu T_{\sigma\mu}-\partial_\sigma T_{\mu\nu}    \\
\mathbf{H}_{\mu\nu\sigma} &= \partial_\mu \mathbf{H}_{\nu\sigma}+\partial_\nu \mathbf{H}_{\sigma\mu}-\partial_\sigma \mathbf{H}_{\mu\nu}.
\end{split}
\end{equation}
Following this construction, we derive the equations of motion via the Euler-Lagrange formalism (see derivation in the Appendix)
\begin{equation} \label{mesonFE}
\begin{split}
(&\Box + m_T^2) T^{\mu \nu} = g_T \bar{\psi} \sigma^{\mu \nu} \psi \\
(&\Box + m_H^2) \mathbf{H}^{\mu \nu}= g_H \bar{\psi} \sigma^{\mu \nu} \boldsymbol{\tau} \psi
\end{split}
\end{equation}

To proceed further, we take the mean-field approximation and limit ourselves to describing spherically symmetric nuclei in their ground state. 
In doing so,we compute the expectation value of the meson fields $\langle T^{\mu \nu} \rangle$, $\langle \mathbf{H}^{\mu \nu} \rangle$. 
The expectation values for the source terms of the RHS of the above equations can be found using the form of the Dirac bilinear shown in \cref{eq:psi}.
We find the nonzero components of each tensor meson are the time-spacelike components ($\mu\nu=0i$), indicating the source terms are, the isoscalar tensor density $\rho_{\textrm{t0}}$ and the isovector tensor density$\rho_{\textrm{t1}}$ \cite{Glendenning:2000,Salinas_thesis}.

\begin{equation} \label{eq:FE_tensor_final}
\begin{split}
    \langle T^{\mu\nu} \rangle = T^{0i} \equiv &T_0(r)\hat{r} \\
    \langle \bm{H}^{\mu\nu} \rangle = H^{0i} \equiv &H_0(r)\hat{r} \\
    g_T \langle \bar{\psi} \sigma^{\mu\nu} \psi \rangle = ig_T \langle \bar{\psi} \vec{\alpha} \psi \rangle \equiv &g_T\rho_{\textrm{t0}}(r) \hat{r} \\
     g_H \langle \bar{\psi} \sigma^{\mu\nu} \bm\tau \psi \rangle = i g_H \langle \bar{\psi} \vec{\alpha} \tau_3 \psi \rangle \equiv &g_H\rho_{\textrm{t1}}(r) \hat{r}
\end{split}
\end{equation}
where we have used $\alpha^i=\gamma^0 \gamma^i$. 
Thus, the equations of motion for each particle reduce to the following form:
\begin{equation}
\begin{split}
\left(m_T^2 + \frac{2}{r^2} - \frac{2}{r} \frac{\partial}{\partial r} - \frac{\partial^2}{\partial r^2} \right) T_0(r) &= g_T\rho_{\textrm{t0}}(r) \\
\left(m_H^2 + \frac{2}{r^2} - \frac{2}{r} \frac{\partial}{\partial r} - \frac{\partial^2}{\partial r^2} \right) H_0(r) &= g_H\rho_{\textrm{t1}}(r).
\end{split}
\label{eq:KG-equations}
\end{equation}

As we would expect for spherical symmetry, the tensor field and source terms only have radial components. For this reason we treat the Proca equation as a scalar differential equation or mathematically project onto $\hat{r}$.
We also note that due to the tensor field being explicitly a vector field, the usual scalar Laplacian is replaced by its vector counterpart, resulting in a Proca equation instead of the usual Klein-Gordon derived for the other mesons.

From here, we can write the Dirac equation for nucleons in the relativistic Hartree approach as follows.
Following the same formalism as Ref.\cite{salinas:2024}, we can write the effective Dirac Hamiltonian as in \cref{eq:hamiltonian}, except now the tensor mesons enter the tensor field in \cref{eq:mesons} ($T^0(r)$) as follows
\begin{equation}
    T(r) = T^0(r) + 2 \mathcal{T}_0(r) + 2\tau_3\mathcal{H}_0(r).
\end{equation}
Here for simplicity, we adopt a shorthand notation for these meson fields as we did for the other mesons,
\begin{center}
$(g_TT_0,g_HH_0)$
    $\rightarrow$
$(\mathcal{T}_0,\mathcal{H}_0)$.
\end{center}

From here, the form of the differential equation for the upper and lower Dirac wavefunctions remains unchanged from \cite{salinas:2024}.
\begin{widetext}
\begin{subequations}
\begin{eqnarray}
  && \left(\frac{d}{dr}+\frac{\kappa}{r}+T(r)\right)g_{n\kappa}(r)-\Big(E_n+M-S(r)-V(r)\Big)f_{n\kappa}(r)=0,\\
  && \left(\frac{d}{dr}-\frac{\kappa}{r}-T(r)\right)f_{n\kappa}(r)+\Big(E_n-M+S(r)-V(r)\Big)g_{n\kappa}(r)=0.
\end{eqnarray}
\end{subequations}
\end{widetext}
The Dirac equation solutions then give rise to point particle densities.
For the tensor mesons, their source densities can be calculated from
\begin{equation} \label{eq:ten_dens}
\begin{split}
    \rho_{\textrm{t}}(r) &= 2 \sum_{n \kappa} \frac{2j_\kappa+1}{4\pi r^2} g_{n\kappa}(r)f_{n \kappa}(r) \\
    \rho_{\textrm{t0}}(r) &= \rho_{\textrm{t,p}}(r) + \rho_{\textrm{t,n}}(r) \\
    \rho_{\textrm{t1}}(r) &= \rho_{\textrm{t,p}}(r) - \rho_{\textrm{t,n}}(r).
\end{split}
\end{equation}
To solve for the tensor meson fields, we fold the source densities in 
\cref{eq:KG-equations} with the static Green's function
\begin{equation}
    D(r, r';m) = \frac{e^{-mr_>}}{4\pi m^2r_>^2 r_<}(1 + mr_>) \left[ \cosh(mr_<) - \frac{\sinh(mr_<)}{mr_<} \right].
    \label{eq:greens_function}
\end{equation}
For example, the isoscalar-tensor meson is solved from the following integral
\begin{eqnarray}
    T_0(r)=g_T\int_0^\infty d^3r' \rho_{t0}(r')D(r,r';m_T)
\end{eqnarray}
A full derivation of this Green's function can be found in the Appendix.
We then iterate solutions of these equations following standard Hartree practices \cite{horowitz:1981} until we achieve self-consistency.
Once self-consistency is achieved, we may calculate nuclear structure properties with these new mesons included, such as, e.g., point proton/neutron densities, binding energies, and nuclear shell structure. 
The nuclear binding energy $E$ is calculated from
\begin{align}
    &\varepsilon_\textrm{tens} = \left(\frac{2}{r^2}\frac{\mathcal{T}_0^2}{g^2_T} - \mathcal{T}_0 \rho_{t0} \right) + \left(\frac{2}{r^2}\frac{\mathcal{H}_0^2}{g^2_H} - \mathcal{H}_0 \rho_{t1} \right)  \\
    &E = E_0 +4\pi \int_0^\infty \varepsilon_\textrm{tens} r^2dr
\end{align}
where $E_0$ is the binding energy without the tensor mesons.

We conclude this section by noting that since the expectation value $\langle \bar{\psi} \boldsymbol{\alpha} \psi \rangle$ vanishes in infinite nuclear matter due to spin-saturation \cite{Glendenning:2000}, the addition of a spin-2 meson plays no direct role in the nuclear matter EOS.
Instead it will have indirect effects through the re-calibration of relativistic density functionals, which we further discuss below.
We also note that this study into these meson's effects only occurs at the Hartree level and further study into their role in meson exchange interactions (or Fock terms) will not be discussed.

\section{Results}
\label{sec:results}
We first note that the addition of these two mesons introduces four new couplings to the RMF Lagrangian, two Yukawa couplings and two masses.
For the meson masses, we choose to fix $m_T=1270$ MeV and $m_H=1320$ MeV so as to match the masses of the $f_2(1270)$ and $a_2(1320)$ mesons, respectively \cite{Suzuki1993,PDG2020}.
As a starting interaction, we use the Chiral-EFT informed RMFGO \cite{reed:2025} interaction and then fix the absolute value of their Yukawa couplings to $|g_T^2|=20$ and $|g_H^2|=100$.
The absolute value is explicitly shown here to account for the possibility of a negative coupling-squared, wherein the particle would be treated as either repulsive ($+g^2$) or attractive ($-g^2$).
To easily distinguish and explore the effects of being repulsive or attractive, we introduce two dimensionless terms $\Gamma_i$ which prepend the Yukawa couplings-squared $(g_T^2,g_H^2)\rightarrow (\Gamma_T |g_T^2|, \Gamma_H|g_H^2|)$.
These $\Gamma_i$ terms only take on values of $\pm1$ and are a phenomenological way to systematically change the effective potential of each particle without changing the Lorentz-covariant structure of the field equations.

\subsection{Impact of Isoscalar-Tensor Meson}
\begin{figure}[htb]
    \centering
    \includegraphics[width=1\linewidth]{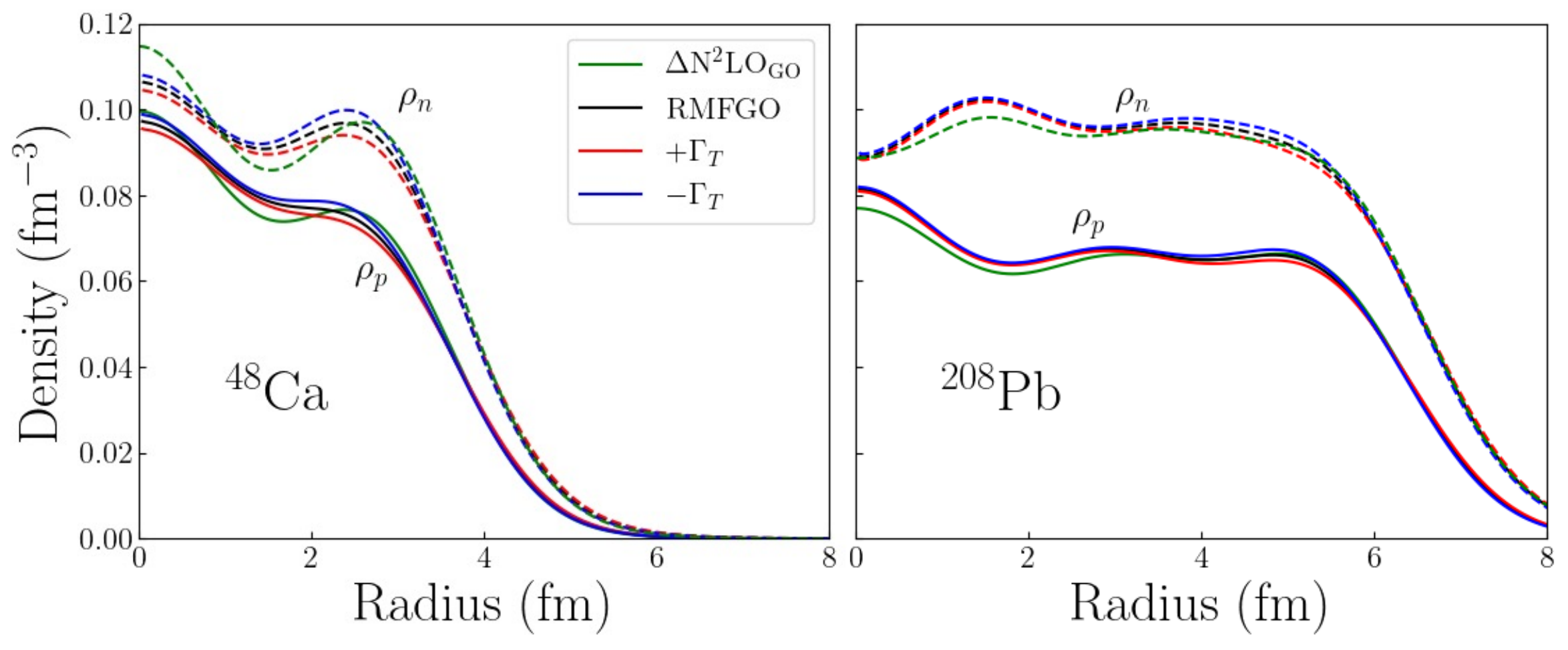}
    \caption{Point particle densities for protons (solid) and neutrons (dashed) in $^{48}$Ca (left) and $^{208}$Pb (right). We show the original RMFGO prediction without tensor mesons as a black line and the $\Delta$N$^2$LO$_{\rm GO}$ interaction which it was original fit in green. The effects of $\pm\Gamma_T$ can be seen best near the nuclear surface in $^{48}$Ca, near $\sim3$ fm.}
    \label{fig:gammaT_ndens}
\end{figure}

\begin{figure}[htb]
    \centering
    \includegraphics[width=1\linewidth]{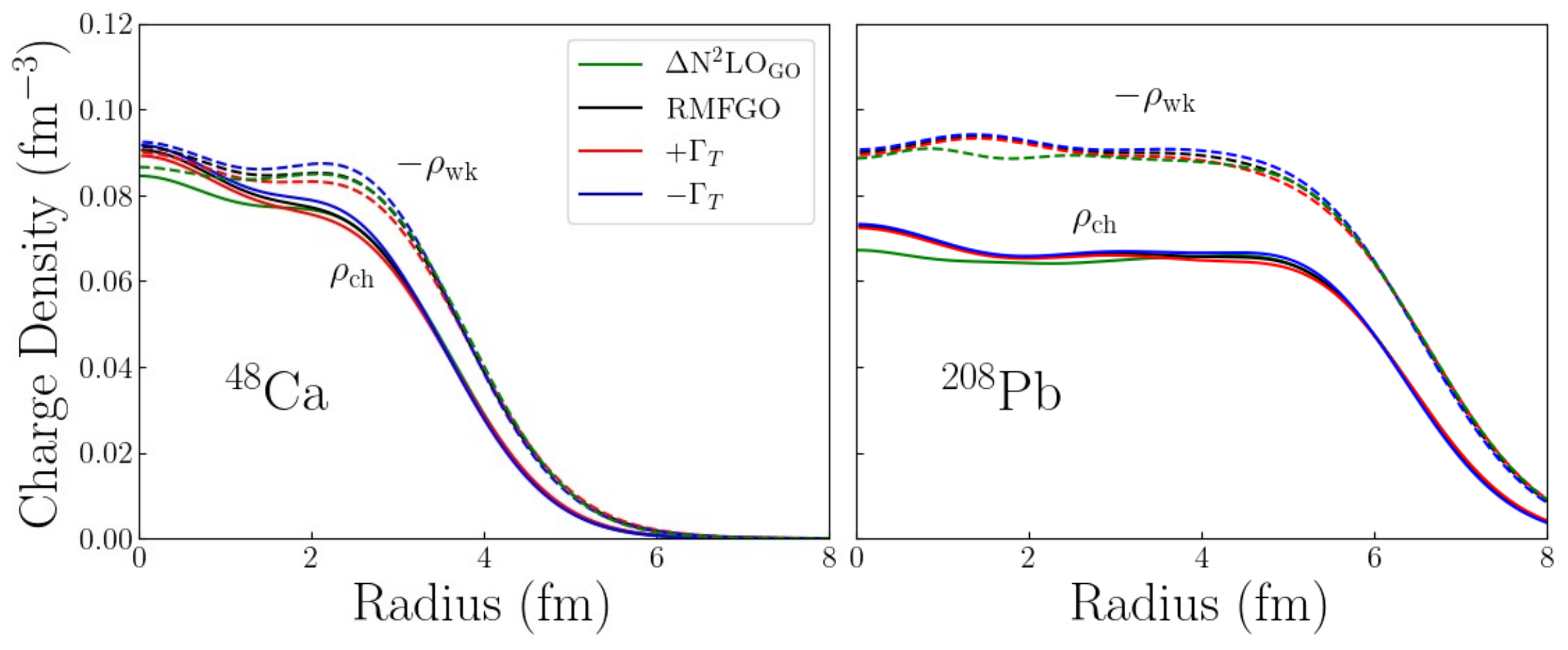}
    \caption{Same as in \cref{fig:gammaT_ndens} but for charge (solid) and minus weak (dashed) densities.}
    \label{fig:gammaT-cdens}
\end{figure}

We begin by showing the impact of the isoscalar-tensor meson on the proton and neutron densities in $^{48}$Ca and $^{208}$Pb in \cref{fig:gammaT_ndens}.
We also plot the point proton and neutron densities calculated using IMSRG(2) \cite{Heinz2021PRC_IMSRG3} from the $\Delta$N$^2$LO$_{\rm GO}$ \cite{Jiang2020PRC_DN2LOGO} interaction of which RMFGO was originally fit in \cite{reed:2025}.
We find that the presence of this meson has the largest impact on the nucleus near its nuclear surface, where we see the greatest change from the RMFGO values depending on the sign of $\Gamma_T$. 
For nuclei with larger surface energies (e.g. $^{48}$Ca) this effect is more pronounced than in $^{208}$Pb where the surface is much smaller.
We also observe substantial changes in the charge and weak densities, shown in \cref{fig:gammaT-cdens}.
In both cases we find that the point densities favor a positive value for $\Gamma_T$ in order to suppress small-r oscillations which are not found in the EFT calculation or in experimental charge densities.

We also show the impact on the physical observables in \cref{tab:obs}.
Firstly, we find that these mesons have the greatest impact on the nuclear binding energy, with $\pm\Gamma_T$ corresponding to a less/more bound nucleus.
This indicates that the isoscalar-tensor meson can be utilized as another coupling to better match the binding energies of nuclei to experimental data.
Furthermore, we see that $\pm\Gamma_T$ can be used to increase/decrease the nuclear radii with about the same sensitivity for each nucleus we present in \cref{tab:obs}.

\begin{figure}
    \centering
    \includegraphics[width=\linewidth]{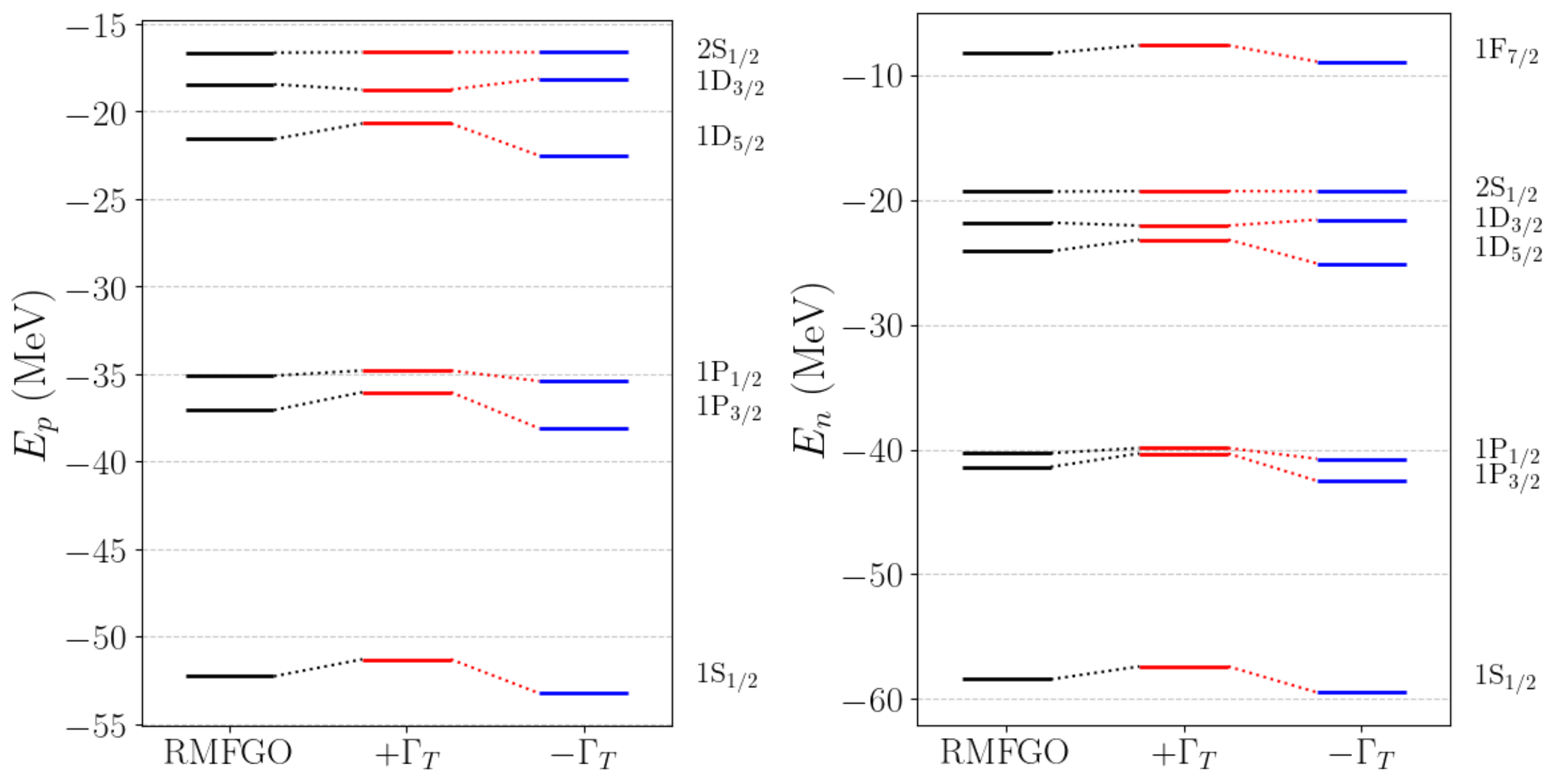}
    \caption{Single particle energies for the different occupied shells in $^{48}$Ca for protons (left) and neutrons (right). The base RMFGO values are shown in the leftmost column as black bars and the effects of changing to $\pm\Gamma_T$ are shown as red and blue bars. Shell state names for each occupied level are shown on the right of each subfigure.}
    \label{fig:gammaT_spectrum}
\end{figure}

It was shown in \cite{Kunjipurayil:2025xss} that introducing too strong of a spin-orbit force can drastically alter the well-structured energy levels in neutron-rich nuclei.
As a check, we provide the occupied levels and eigenenergies of the various proton and neutron shells in $^{48}$Ca in \cref{fig:gammaT_spectrum}.
Here we observe that this meson has the greatest impact on the $1P_{3/2}$ and $1D_{5/2}$ shell states in both protons and neutrons.
There is a small impact on the other occupied shell states as well, but we do not observe any level crossings as in \cite{Kunjipurayil:2025xss}.

\begin{table*}
    \renewcommand{\arraystretch}{1.3}
    \begin{ruledtabular}
    \begin{tabular}{l c c c c c c c c c c c c c}
    Model & $m_s$ & $g_s^2$ & $g_v^2$ & $g_\rho^2$ & $g_\delta^2$ & $\kappa$ & $\lambda$ & $\zeta$ & $\Lambda_v$ &   $f_v$ & $f_\rho$\\\hline\hline
    RMFGO& $ 523.2300 $ & $ 129.8895 $ & $ 207.3669 $ & $ 187.2445 $ & $ 87.2318 $ & $ 2.0458 $ & $ 0.0245 $ & $ 0.0756 $ & $ 0.0198 $  &  $ 11.6757 $ & $ -107.0218 $\\
    FSUGold2(L90) & 497.479 & 108.0943 & 183.7893 & 91.1864 & 0.0 & 3.0029 & -0.000533 & 0.0256 & 0.009393 & 0.0 & 0.0\\\hline

    
    \end{tabular}
    \end{ruledtabular}
    \caption{Meson field couplings for the RMFGO and FSUGold2(L90)  interactions used in the text. Note that the scalar meson mass $m_s$ is given in units of MeV. We fix the masses of the other mesons to their experimental values of $m_v=782.5$ MeV, $m_\rho=763$ MeV, and $m_\delta=980$ MeV \cite{PDG2020}.}
    \label{tab:yukawa_couplings}
\end{table*}

\begin{table}
\centering
\begin{tabular}{| c | c | c | c | c | c | c | c |}
\hline
 & Observable & RMFGO & Experiment & $+\Gamma_T$ & $-\Gamma_T$ & $+\Gamma_H$ & $-\Gamma_H$ \\
\hline
 & $R_p$ (fm) & 3.29 & -- & 3.31 & 3.28 & 3.29 & 3.29 \\
 & $R_n$ (fm) & 3.23 & -- & 3.25 & 3.22 & 3.23 & 3.23 \\
$^{40}$Ca & $R_n-R_p$ (fm) & -0.06 & -- & -0.06 & -0.06 & -0.06 & -0.06 \\
 & $E/A$ (MeV) & -8.61 & -8.55 & -8.51 & -8.71 & -8.61 & -8.60 \\
 & $R_{\rm ch}$ (fm) & 3.38 & 3.48 & 3.40 & 3.36 & 3.38 & 3.38 \\
 & $R_{\rm wk}$ (fm) & 3.32 & -- & 3.34 & 3.30 & 3.32 & 3.32 \\
\botrule
 & $R_p$ (fm) & 3.37 & -- & 3.40 & 3.34 & 3.37 & 3.37 \\
 & $R_n$ (fm) & 3.53 & -- & 3.56 & 3.49 & 3.57 & 3.48 \\
$^{48}$Ca & $R_n-R_p$ (fm) & 0.16 & $0.12\pm.03$\cite{CREX} & 0.17 & 0.14 & 0.20 & 0.11 \\
 & $E/A$ (MeV) & -8.42 & -8.67 & -8.25 & -8.60 & -8.31 & -8.53 \\
 & $R_{\rm ch}$ (fm) & 3.43 & 3.48 & 3.46 & 3.40 & 3.43 & 3.43 \\
 & $R_{\rm wk}$ (fm) & 3.64 & $3.64\pm.03$ \cite{CREX}& 3.67 & 3.60 & 3.68 & 3.59 \\
 \botrule
  & $R_p$ (fm) & 5.36 & -- & 5.39 & 5.34 & 5.36 & 5.36 \\
 & $R_n$ (fm) & 5.52 & -- & 5.55 & 5.50 & 5.53 & 5.52 \\
$^{208}$Pb & $R_n-R_p$ (fm) & 0.16 & $0.28\pm.07$\cite{PREX2} & 0.16 & 0.16 & 0.17 & 0.16 \\
 & $E/A$ (MeV) & -7.95 & -7.81 & -7.85 & -8.06 & -7.94 & -7.97 \\
 & $R_{\rm ch}$ (fm) & 5.40 & 5.50 & 5.43 & 5.38 & 5.40 & 5.40 \\
 & $R_{\rm wk}$ (fm) & 5.59 & $5.60\pm.07$\cite{PREX2}& 5.62 & 5.57 & 5.60 & 5.59 \\
 \hline
\end{tabular}
\caption{Properties of select nuclei for the tensor mesons. We show both the results of using $\pm\Gamma_i$ for both nuclei compared to the RMFGO interaction. We use the experimental binding energies from Ref. \cite{Kondev_2021} and charge radii from Ref. \cite{Angeli:2013}.}
\label{tab:obs}
\end{table}

\subsection{Impact of Isovector-Tensor Meson}
\begin{figure}[htb]
    \centering
    \includegraphics[width=1\linewidth]{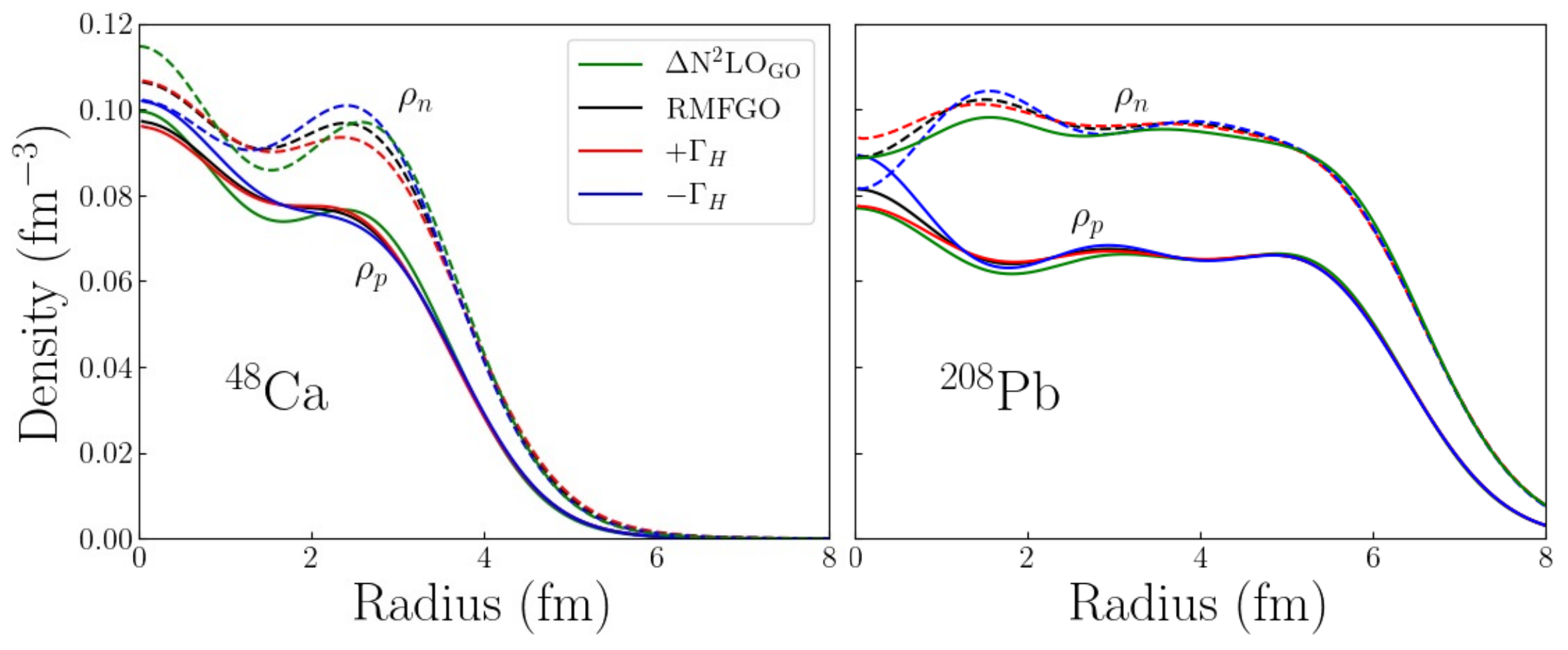}
    \caption{Same as in \cref{fig:gammaT_ndens} but for changing $\Gamma_H$.}
    \label{fig:gammaH_ndens}
\end{figure}

\begin{figure}[htb]
    \centering
    \includegraphics[width=1\linewidth]{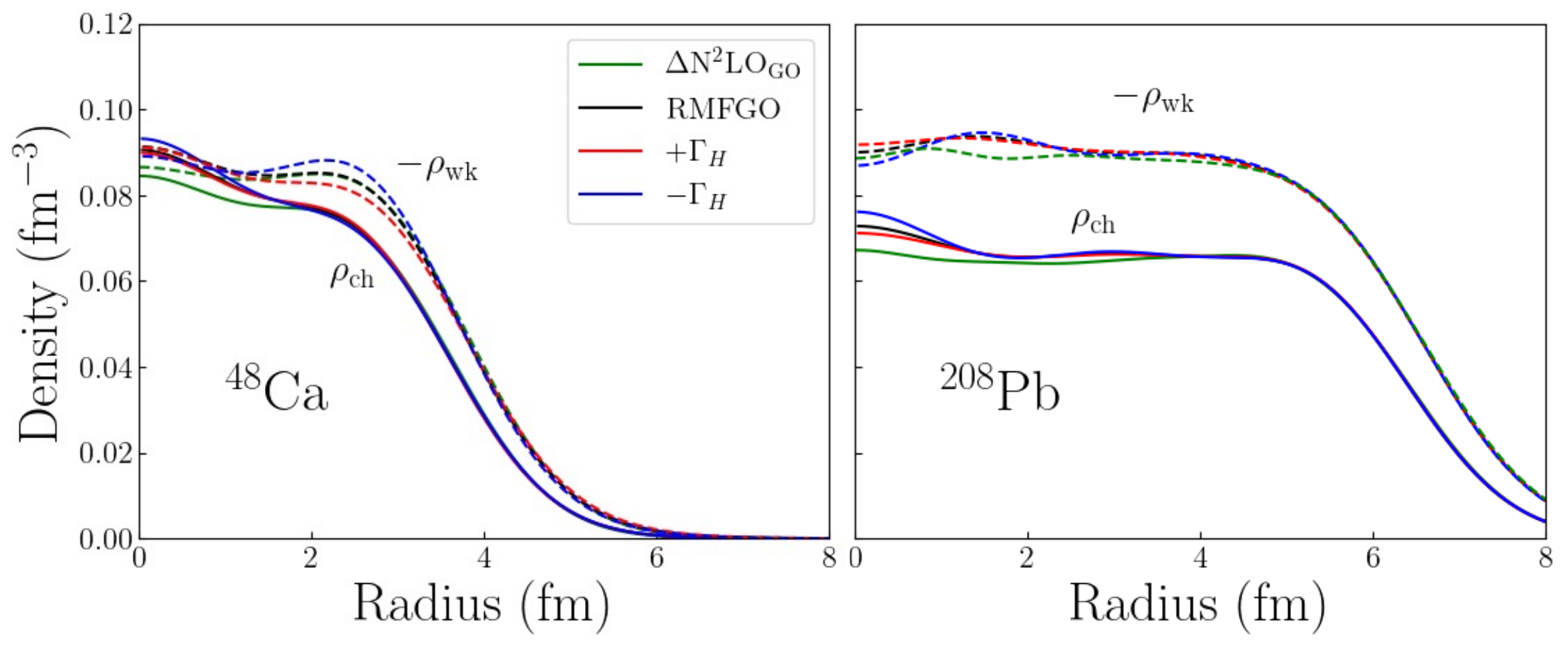}
    \caption{Same as in \cref{fig:gammaT-cdens} but for changing $\Gamma_H$.}
    \label{fig:gammaH-cdens}
\end{figure}

As in the previous section, we now discuss the impact of the isovector-tensor meson.
Similar to \cref{fig:gammaT_ndens} and \cref{fig:gammaT-cdens}, we show the nucleonic and charge densities of $^{48}$Ca and $^{208}$Pb in \cref{fig:gammaH_ndens} and \cref{fig:gammaH-cdens}, respectively.
We observe some of the same features as with the isoscalar-tensor meson, with the $\pm\Gamma_H$ corresponding to increasing/decreasing the densities near the nuclear surface.
One noticeable difference is that the isovector-tensor meson greatly affects the neutron density compared to the proton density, a feature to be expected from an isospin-dependent meson field.
This could prove powerful as providing an additional degree of freedom to isovector forces in nuclei, which could further help remedy the PREX-CREX ``dilemma''\cite{reed:2025,Kunjipurayil:2025xss}.

\begin{figure}
    \centering
    \includegraphics[width=\linewidth]{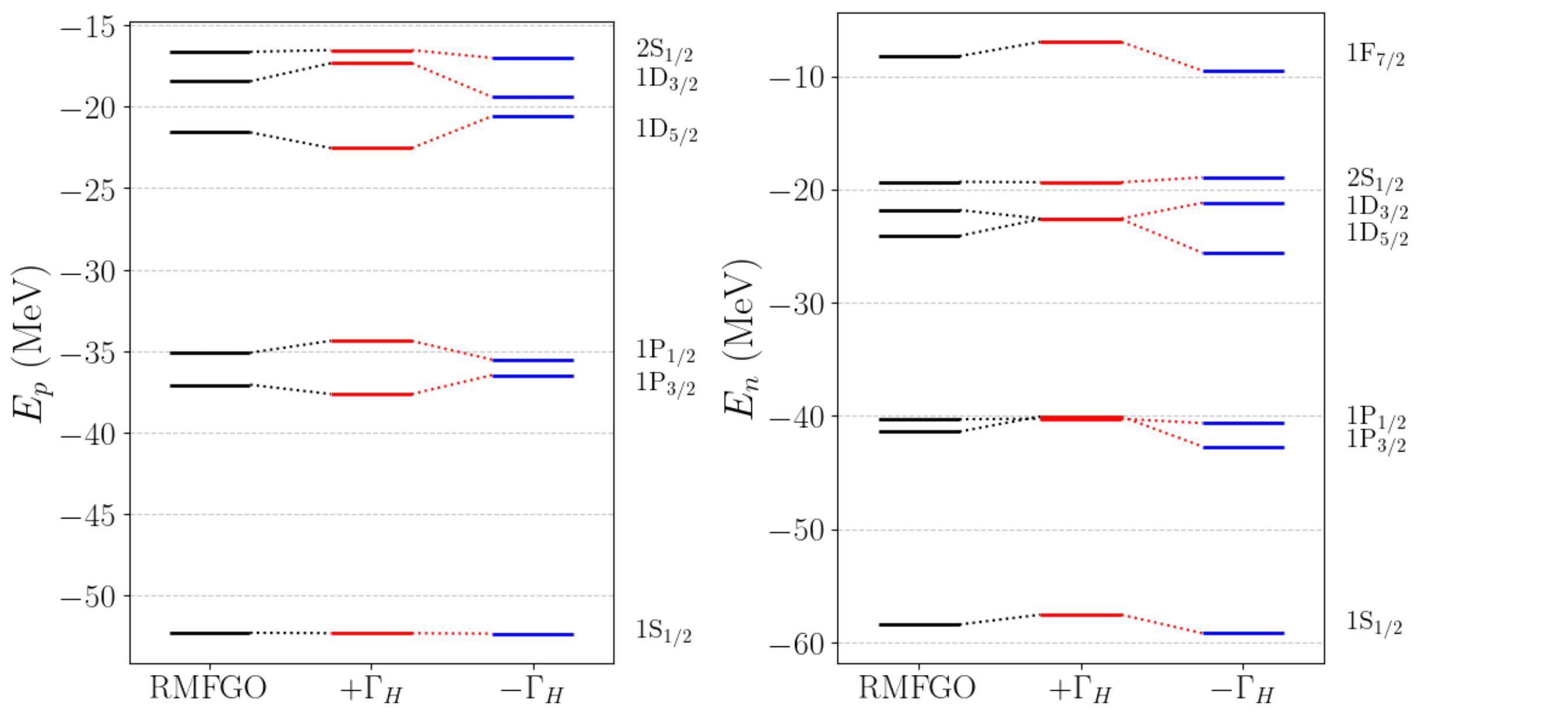}
    \caption{Same as \cref{fig:gammaT_spectrum} but for changing $\Gamma_H$ values.}
    \label{fig:gammaH_spectrum}
\end{figure}

The effect of this meson on nuclear observables can be seen in the same \cref{tab:obs}.
Here, we see that this meson has little effect on the symmetric $^{40}$Ca nucleus, but has a great impact on the properties of the neutron-rich $^{48}$Ca and $^{208}$Pb nuclei.
We attribute this to the meson's isovector nature.
Compared to the isoscalar-tensor meson, however, we find that this meson has a much weaker effect on nuclear structure observables for both $\pm\Gamma_H$, with the exception of the neutron skin.
We do see a significant difference between the two mesons' impact in the nuclear energy spectra in \cref{fig:gammaH_spectrum}, where we plot the energy of occupied shells for protons and neutrons as in \cref{fig:gammaT_spectrum}.
Here we find similar effects for the proton spectrum but a significant difference in the neutron spectrum.
For $+\Gamma_H$ there is the potential for a level crossing for the $1P$ and $1D$ shell pairs which we are hoping to avoid.
This is indicative that to preserve the well-ordered shell structure in $^{48}$Ca that $\Gamma_H$ should be negative.
Therefore, we conclude that choosing $+\Gamma_T$ ($g_T^2>0$) and $-\Gamma_H$ ($g_H^2<0$) provides the effect on nuclear structure that we seek without sacrificing reliability with well-known nuclear shell orderings.

\begin{figure}[h]
    \centering
    \includegraphics[width=0.95\linewidth]{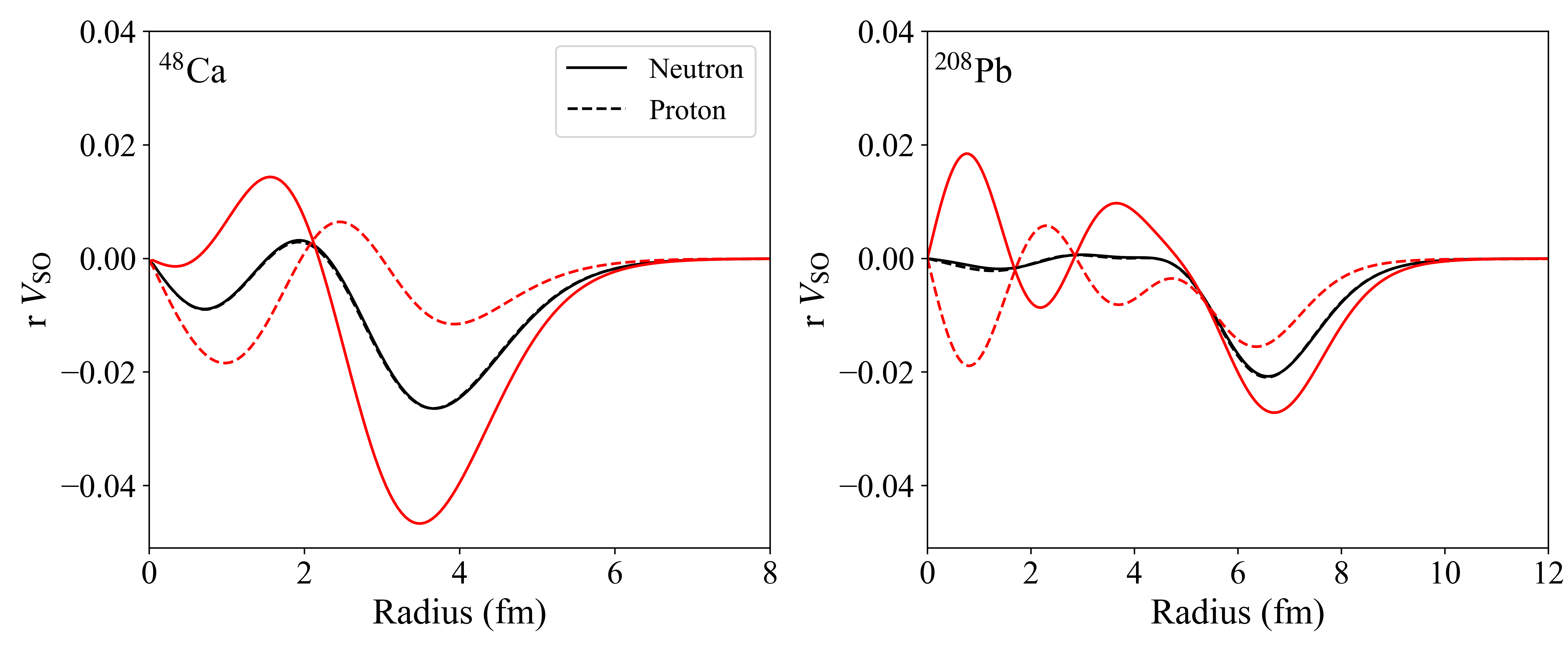}
    \caption{The neutron and proton spin-orbit potentials are shown for the FSUGold2(L90) model (black) with $^{48}$Ca (left) and $^{208}$Pb (right). The red lines show the same spin-oribt potential of FSUGold2(L90), but with the isovector tensor coupling ($g_H^2 = -400$). Here $rV_\textrm{SO}$ is dimensionless.}
    \label{fig:SO_potential}
\end{figure}

Furthermore, we conclude this section by investigating the direct effect on the isovector spin-orbit potential.
Firstly, we present the spin-orbit potential of the FSUGold2(L90) \cite{Reed_2021p2} interaction in both $^{48}$Ca and $^{208}$Pb in \cref{fig:SO_potential}.
We also overplot the same interaction but with $g_H^2=-400$ to show the large impact this coupling has.
Firstly, we observe that the spin-orbit potential in the bare interaction has a nearly identical behavior for both protons and neutrons.
When adding in the H-meson, we see significant splitting where the neutron spin-orbit density peaks at the nuclear surface in both nuclei.
In the interior densities below the surface of $^{208}$Pb, we see that the proton and neutron densities mirror each other, which is expected following the work of \cite{Kunjipurayil:2025xss} which showed a similar behavior for an enhanced spin-orbit potential.
Primarily, we find that the inclusion of the H-meson here leads to an isovector dominance in the spin-orbit potential, contrary to what is seen in traditional RMF models wherein the isoscalar contribution is much larger.

\begin{figure}
    \centering
    \includegraphics[width=\linewidth]{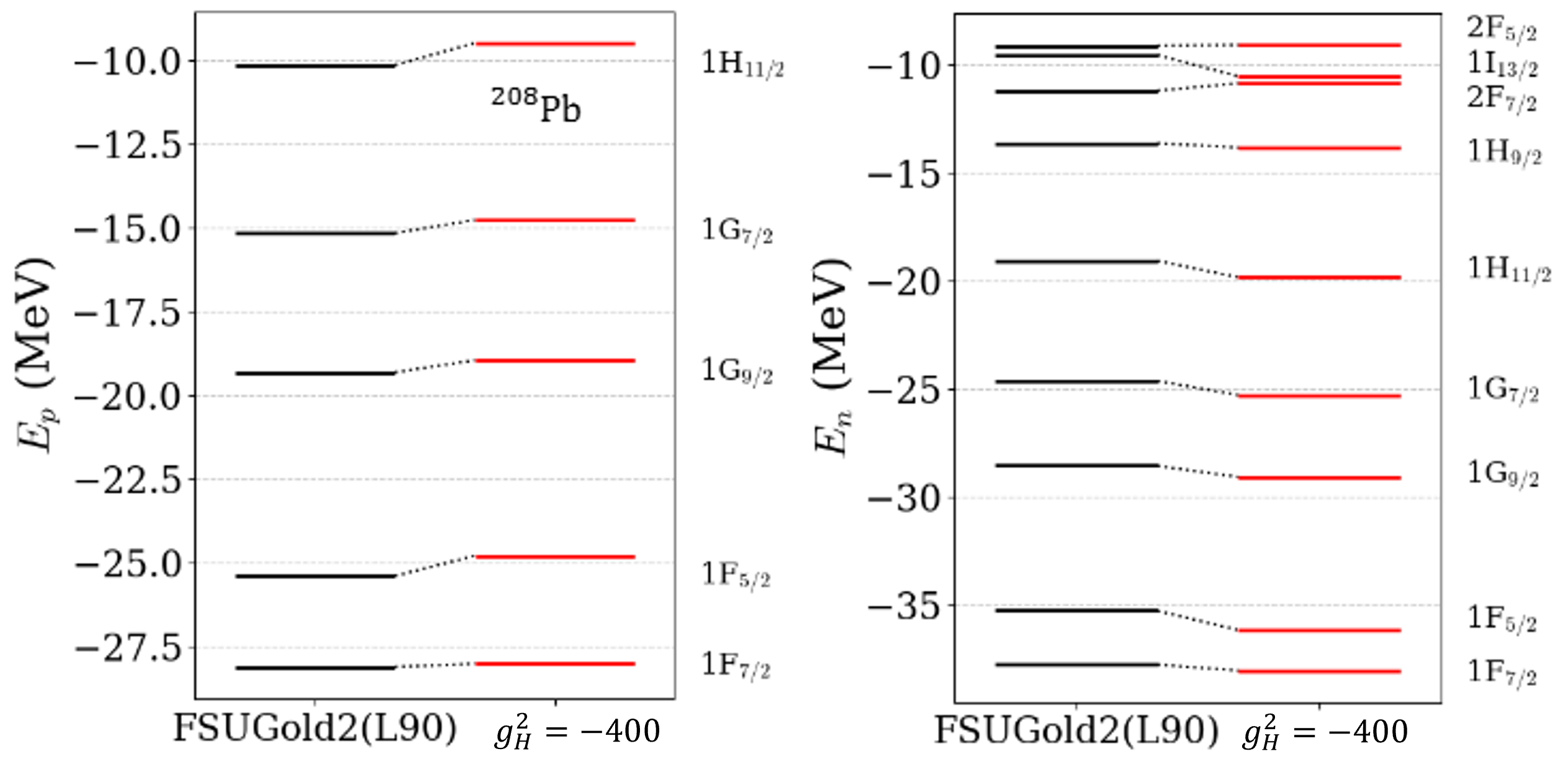}
    \caption{Single particle energies for occupied $l\geq3$ shells in $^{208}$Pb for the bare FSUGold2(L90) RMF interaction (black) and with the large H-meson coupling added (red).}
    \label{fig:spectrum_lead}
\end{figure}

It was shown that a consequence of such an enhancement is the observation of level crossings in the shell structure of heavy nuclei such as $^{208}$Pb \cite{Kunjipurayil:2025xss}.
To further probe the impact of such an enhancement to the isovector spin-orbit potential, we also plot the level diagrams for $^{208}$Pb using FSUGold2(L90) both with and without the H-meson in \cref{fig:spectrum_lead}.
To investigate possible level-crossings found for states with larger angular momentum, we show the occupied states with $l\geq3$.
As previously seen within $^{48}$Ca, we do not find any level crossings even for such large Yukawa couplings for the H-meson.
This is promising that these mesons could be used as an efficient way to enhance the isovector spin-orbit sector of the RMF Lagrangian while also mitigating the level crossing of the $1I_{13/2}$ and $2F_{7/2}$ orbitals, one of the bigger issues seen in models with this type of enhanced isovector spin-orbit interaction \cite{Kunjipurayil:2025xss}. 

\section{Discussion}
\label{sec:discussion}
We first wish to discuss the nature of these tensor mesons in the grand scheme of covariant field theory.
The study of relativistic spin-2 particles in literature largely has been explored in the regime of gravitons, the massless spin-2 mediator of the gravitational force.
For \textit{massive} spin-2 particles, they have been much less studied in the context of effective field theory and especially in the context of nuclear forces.
An additional structural limitation of our study is in the construction of the coupling to the nucleonic fields through the $\sigma^{\mu\nu}$ tensor.
Inherently this forces these mesons to be antisymmetric and traceless, an uncommon feature of mesons and in the study of spin-2 bosons.
Thus, we must stress that these mesons inherently are not the $f_2(1270)$ and $a_2(1320)$ particles but instead are an effective field used to mimic some of the stronger isovector spin-orbit interactions implied by the PREX/CREX experiments and the work of \cite{Yue:2024,Kunjipurayil:2025xss}.
Much in a similar way that the scalar $\sigma$ meson, although often compared to the $f_0(500)$ particle, is distinct in its inclusion to the RMF Lagrangian.

An important goal of including these tensor mesons is to expand the spin-orbit sector of the RMF Lagrangian without relying on the other mesons found in the Lagrangian.
Previously, these tensor interactions were only included through the $\omega$ and $\rho$ meson tensor terms in \cref{eq:tensor_couplings} in RMF theory.
As stated before, this inherently ties their effects and behavior in nuclei to the meson fields which are present in both nuclei and nuclear matter calculations.
Thus, the new tensor mesons allow for an independent tuning to the spin-orbit potential in finite nuclei.
Additionally, we include an attractive isovector-tensor meson $\bm{H}_{\mu\nu}$ and a repulsive isoscalar-tensor meson $T_{\mu\nu}$ field.

The repulsive meson contribution implies that the Yukawa-like coupling-squared is positive whereas the attractive meson is \textit{negative}.
This requires that the bare coupling, $\sqrt{g_H^2}=g_H$, is imaginary, implying that the underlying field is imaginary as well.
This is merely a phenomenological choice we make as the underlying physical observables, these being densities and particle energies, for example, are real-defined.
It is interesting though to point out that this particle differs greatly from the other mesons in the standard RMF Lagrangian in this parameterization.

We have presented here the effects on finite nuclei of adding two exotic meson fields to the RMF Lagrangian in the Hartree approximation.
These spin-2 particles are unique in their inclusion as they contribute to the nuclear structure distinctly from any other interaction currently presented in state-of-the-art Lagrangians.
Typically in DFT Hartree calculations, the tensor forces are obtained through meson exchange potentials, in particular the two-body pion exchange seen in relativistic Hartree-Fock calculations and in EFT many-body calculations.
Obtaining tensor forces from pion exchange or contact interactions at the Hartree level is difficult to parameterize, as they do not contribute to spin-saturated nuclei in the mean-field approximation.
This requires that the effects of the tensor force be absorbed within other effective terms which we do here through the tensor mesons.

Unique to our description presented here, the tensor mesons act as mediators for the spin-orbit tensor force seen in Skyrme parameterizations \cite{Yue:2024}, distinct from the form of tensor interactions in EFT parameterizations.
However unlike in EFT calculations, these tensor mesons do not contribute to nuclear matter and thus are only impactful in finite nuclei.
It is possible that their inclusion to nuclear matter via Hartree-Fock calculations of matter may be impactful and would be an excellent follow-up to this study.
Their propagator, if parameterized for illustration as similar to the propagator for the other mesons in RMF theory, only becomes nontrivial for very large momentum transfers since their masses are so much larger than any other particle in the standard RMF Lagrangian.
Thus, we predict that their contributions be small.

Lastly, we wish to discuss what these mesons mean in the context of the PREX/CREX ``dilemma".
This is to say, how can the neutron skins of $^{208}$Pb and $^{48}$Ca be so far off from the trend expected by nuclear models?
Indeed since the announcement of the results of the PREX-2 and CREX experiments there has been quite the stir in the nuclear structure community to reconcile these two results to within 90\% confidence, a feat which has proven to be quite difficult using traditional DFT methods.
It has been shown that by enhancing the isovector spin-orbit sector of DFT models, one can produce a large neutron skin in $^{208}$Pb and a systematically smaller skin in $^{48}$Ca \cite{Reed:2023,Yue:2024,Kunjipurayil:2025xss}.
This work introduces two mesons which, in the non-relativistic limit, do contribute to the spin-orbit potential directly.
However, since the theoretical implications of both experiments have recently been called into question \cite{roca-maza:2025,jakubassa:2025}, we merely present that these mesons provide an additional method of varying isovector observables in atomic nuclei.

One avenue of systematic improvements to addressing this ``dilemma" beyond improvements of the nuclear Hamiltonian would be in detailing additional corrections that are important to the theoretical calculation of the parity-violating asymmetry.
Notably, the theory calculations provided in the analysis of both PREX and CREX include tree-level plus coulomb distortions \cite{Horowitz:1998vv} and $\gamma Z$-box corrections \cite{Gorchtein:2008px} as the only radiative corrections to the scattering cross-section.
As was shown in \cite{roca-maza:2025}, additional corrections from QED also are important and may lead to unphysical inference of the neutron-skin in $^{208}$Pb.
However, the corrections from QED alone are incomplete \cite{Reed:2026_comment} and thus, more study into the radiative corrections to the parity-violating asymmetry may be more fruitful to resolving the so-called ``dilemma.''
Indeed nuclear dispersion corrections to the weak and coulomb scattering amplitudes have been shown to be important in estimating the theoretical uncertainty budget of both PREX and CREX \cite{jakubassa:2025}.

The resolution of this ``dilemma'' continues to be important in the study of neutron-rich nuclei and is invaluable to the interpretation of the future Mainz Radius Experiment (MREX) \cite{MREX}.
MREX will provide a third measurement of the parity-violating asymmetry in $^{208}$Pb which will provide an additional measurement of the neutron skin of $^{208}$Pb at a much smaller energy, $\approx50-150$ MeV, as opposed to PREX, $\approx1$ GeV.
Additionally, it is expected that the systematic uncertainties will be nearly \textit{half} that of PREX2, thus increasing the precision by a factor of two.
For this reason, theoretical uncertainties need to be documented and improved upon wherever possible to ensure the success of MREX and the possibility of future measurements of the neutron skin in neutron-rich nuclei.

\section{Conclusion}
\label{sec:conclusions}
We have introduced a new class of spin-2 massive mesons to the nonlinear relativistic mean field theory Lagrangian.
In doing so, we show that these mesons indeed meaningfully contribute to the properties of finite nuclei and uniquely contribute to the spin-orbit sector of the theory.
We also find that despite enhancing the spin-orbit potential greatly, these mesons do not disrupt the well-ordered structure of nuclear shells in finite nuclei.
Due to the tensor-like structure these particles interact with nucleons in, they do not contribute to the energy density of infinite nuclear matter and thus \textit{only} affect the properties of nuclei at the Hartree level.
While it is promising that these mesons allow for potentially increased sophistication of covariant density functional theory, the PREX+CREX ``dilemma'' continues to elude nuclear models.
However, this ``dilemma'' may be mitigated instead by accurate calculation of the parity-violating asymmetry with radiative corrections correctly implemented beyond the tree-level. 

\begin{acknowledgements}
We would like to thank Ingo Tews, Ryan Curry, and Joe Carlson for useful discussions regarding the tensor interaction. 
We would also like to thank Jorge Piekarewicz and C.J. Horowitz for assistance in the derivation of the field equations and for invaluable guidance throughout the project. 
This work was supported by the U.S. Department of Energy through the Los Alamos National Laboratory. 
Los Alamos National Laboratory is operated by Triad National Security, LLC, for the National Nuclear Security Administration of U.S. Department of Energy (Contract No.~89233218CNA000001).
B.T.R. was also supported in part by the U.S. Department of Energy, Office of Science, Office of Nuclear Physics program under Award Number DE-SCL0000015 and by the Laboratory Directed Research and Development Program of Los Alamos National Laboratory under project numbers 20230785PRD1 and 20230315ER. 
This work was also performed under the auspices of the U.S. Department of Energy by Lawrence Livermore National Laboratory under Contract DE-AC52-07NA27344.
\end{acknowledgements}

\appendix
\section*{Appendix}
Here we detail the derivation of all relevant equations necessary for the Hartree approach we use in the main paper.
For this section, we use a general rank-2 tensor $H_{\mu\nu}$ to represent the structure of the two tensor mesons used in the main paper.
Since the equations only differ in their construction with an isospin matrix $\bm{\tau}$, we only write one set of equations in each section for simplicity.

\section{Euler-Lagrange Equation}
We first begin by proposing the following general addition to the RMF Lagrangian, where $\Lambda$ is an arbitrary constant to be fixed/determined later in order to recover the familiar field equations. 
\begin{equation}
    \mathcal{L}_{\rm T} = \frac{\Lambda}{6} H_{\mu \nu \sigma} H^{\mu \nu \sigma} - \frac{1}{2} m_{H}^2 H_{\mu \nu} H^{\mu \nu} + g_H \bar{\psi} \sigma^{\mu \nu} H_{\mu \nu} \psi
\end{equation}
where the spin-2 strength tensor is also written in a general form with unknown constants $a,b,$ and $c$,
\begin{equation}
    H_{\mu \nu \sigma} = a \partial_\mu H_{\nu \sigma} + b \partial_\nu H_{\sigma \mu} + c \partial_\sigma H_{\mu \nu} 
\end{equation}
Note that this form of the strength tensor is adapted and simplified from \cite{Dalmazi:2013cna,Koenigstein:2015asa} to match the symmetries of the tensor mesons.
Additionally, we allow for the Yukawa-like coupling $g_H$ to take on \textit{negative} values which may allow for different physical implications without breaking Lorentz symmetry.
We then take the first variation to derive the Euler-Lagrange equations,
\begin{equation}
\begin{split}
\partial_\alpha \frac{\partial \mathcal{L}}{\partial (\partial_{\alpha} H_{\lambda \gamma})} &= \frac{\partial \mathcal{L}}{\partial H_{\lambda \gamma}} \\
\frac{\partial \mathcal{L}}{\partial H_{\lambda \gamma}} &= g_H \bar{\psi} \sigma^{\lambda \gamma} \psi - m_H^2 H^{\lambda \gamma} \\
\partial_\alpha \frac{\partial \mathcal{L}}{\partial (\partial_{\alpha} H_{\lambda \gamma})} &= \frac{\Lambda}{3} \partial_\alpha[a H^{\alpha \lambda \gamma} + b H^{\gamma \alpha \lambda} + c H^{\lambda \gamma \alpha}]
\end{split}
\end{equation}
arriving at the following general equation of motion
\begin{equation}
\frac{\Lambda}{3} \partial_\alpha[\partial^\alpha H^{\lambda \gamma} (a^2 + b^2 + c^2) + (\partial^\lambda H^{\gamma \alpha} + \partial^\gamma H^{\alpha \lambda}) (ab +bc + ac)] = g_H \bar{\psi} \sigma^{\lambda \gamma} \psi - m_H^2 H^{\lambda \gamma}
\label{eq:EOM_full}.
\end{equation}
With this equation, it becomes clear that the structure of the tensor mesons will need to follow from the nonzero components of $\sigma^{\mu \nu}$.
By its construction, $\sigma^{\mu \nu}=\frac{i}{2}[\gamma^\mu,\gamma^\nu]$ is both traceless and antisymmetric due to the algebra of the gamma matrices.
From this, it follows that $H_{\mu \nu}$ must be an anti-symmetric tensor in order for the Yukawa interaction, $\sigma^{\mu \nu} H_{\mu \nu}$, not to vanish.
Thus we have the following identity, $$H^{\mu \nu} = -H^{\nu \mu}$$, which will prove to be useful later. To make further progress, one must now evaluate the nonzero components of the source density term $g_H \bar{\psi} \sigma^{\lambda \gamma} \psi$.

\section{Source Density and Non-vanishing Components}
Because $\sigma^{\mu\nu}$ is antisymmetric and traceless, we only have to solve for two terms to find all contributions in the mean-field approximation.
First we start with the source term
\begin{eqnarray}
    g_H \bar{\psi} \sigma^{\lambda \gamma} \psi \rightarrow g_H \bar{\psi}
    \begin{pmatrix}
        \sigma^{0i}\\
        \sigma^{ij}
    \end{pmatrix}
    \psi
\end{eqnarray}
and note that the form of the nucleon wavefunctions are
\begin{eqnarray}
    \psi(x) = \frac{1}{r} \begin{pmatrix}
        g_{n \kappa}(r)\mathcal{Y}_{\kappa m}(\mathbf{\hat{r}})\\
        if_{n \kappa}(r)\mathcal{Y}_{-\kappa m}(\mathbf{\hat{r}})
    \end{pmatrix}.
\end{eqnarray}
where $\mathcal{Y}_{\kappa m}(\mathbf{\hat{r}})$ are the tensor spherical harmonics.
For the time-spacelike components, it can be readily shown that $\langle \bar{\psi} \sigma^{0i} \psi \rangle \equiv \rho_{\textrm{t}}(r) \hat{r}$ \cite{salinas:2024}.
For the off-diagonal space-spacelike components, due to spherical and spin symmetry, this value is zero when taking the mean-field approximation, i.e. $\langle \bar{\psi} \sigma^{ij} \psi \rangle \equiv 0$.
Thus, the only contributing terms to the equations of motion are the time-spacelike components of $H_{\mu\nu}$.

\section{Free Field Equations}
If we restrict ourselves to $a,b,c$ being unimodular, that is $(\pm1)$, then it follows that $a^2 + b^2 + c^2 =3$ and $ab+bc+ac = {-1, 3}$.
Therefore we have the following two possible equations of motion for the tensor meson LHS of \cref{eq:EOM_full}:

\begin{equation}
\begin{split}
&(\textrm{Case -1}) \quad \frac{\Lambda}{3} \partial_\alpha[3 \partial^\alpha H^{0i} - \partial^0 H^{i \alpha} - \partial^i H^{\alpha 0}] = g_H\bar{\psi} \sigma^{0i} \psi - m_H^2 H^{0i} \\
&(\textrm{Case +3}) \quad \Lambda \partial_\alpha[ \partial^\alpha H^{0i} + \partial^0 H^{i \alpha} + \partial^i H^{\alpha 0}] = g_H \bar{\psi} \sigma^{0i} \psi - m_H^2 H^{0i}.
\end{split}
\end{equation}
Since we are working within the static limit, all time derivatives vanish ($\partial_\alpha=-\partial_j$).

For the (-1) case we may then simplify to,
\begin{equation}
\begin{split}
& \frac{\Lambda}{3} \partial_j[3 \partial^j H^{0i} - \partial^i H^{j 0}] + m_H^2 H^{0i} = g_H \bar{\psi} \sigma^{0i} \psi \\
& \Lambda [\partial_j \partial^j H^{0i} - \frac{1}{3}\partial_j \partial^i H^{j 0}] + m_H^2 H^{0i} = g_H \bar{\psi} \sigma^{0i} \psi
\end{split}
\end{equation}
Using the antisymmetric property of $H^{\mu\nu}$ we rewrite the above equation as,
\begin{equation}
\Lambda [\partial_j \partial^j H^{0i} + \frac{1}{3}\partial_j \partial^i H^{0 j}] + m_H^2 H^{0i} = g_H \bar{\psi} \sigma^{0i} \psi
\end{equation}
We also recast the tensor field into its antisymmetric and symmetric parts, which in the spherically symmetric limit takes the form,
\begin{equation}
H^{\mu \nu} = (g^{\mu i} g^{\nu 0} - g^{\mu 0} g^{\nu i})H_0(r) \hat{r}_i
\end{equation}
Thus, $H^{0i} = H_0(r) \hat{\mathbf{r}}$ leading to the following mean field equation:
\begin{equation}
\Lambda [-\nabla^2 (H_0(r) \hat{\mathbf{r}}) - \frac{1}{3} \vec{\nabla}(\vec{\nabla} \cdot  H_0(r) \hat{\mathbf{r}})] +  m_H^2 H_0(r) \hat{\mathbf{r}} = g_H \langle \bar{\psi} \sigma^{0i} \psi \rangle
\end{equation}
Using the identity $\vec{\nabla}(\vec{\nabla} \cdot  \vec{A}) = \nabla^2 \vec{A} + \vec{\nabla} \times \vec{\nabla} \times \vec{A}$, and noting that the cross product term vanishes due to the spherical symmetry we get:
\begin{equation}
\left[ -\frac{4 \Lambda}{3}\nabla^2 + m_H^2 \right]H_0(r) \hat{\mathbf{r}} = g_H \langle \bar{\psi} \sigma^{0i} \psi \rangle
\end{equation}
For $\Lambda = 3/4$, we can recover the static Proca equation. 

One important thing to note is that the usual Laplacian operator is now replaced by the vector Laplacian giving rise to a different Green's function than the typical one used in this kind of RMF.
This combined with the results for the value of the source term in \cref{eq:EOM_full} leads to the following radial differential equation for the massive spin-2 meson 
\begin{equation}
\boxed{
\left(m_H^2 + \frac{2}{r^2} - \frac{2}{r} \frac{\partial}{\partial r} - \frac{\partial^2}{\partial r^2} \right)H_0(r) = g_H \rho_t(r)}
\label{eq:proca-eq}
\end{equation}.

Similarly for Case +3, we can simplify the field equation to
\begin{equation}
\begin{split}
    \Lambda [\partial_j \partial^j H^{0i} + \partial_j \partial^i H^{j 0}] + m_H^2 H^{0i} = g_H \bar{\psi} \sigma^{0i} \psi \\
    \Lambda [\partial_j \partial^j H^{0i} - \partial^i \partial_j H^{0 j}] + m_H^2 H^{0i} = g_H \bar{\psi} \sigma^{0i} \psi \\
    \Lambda [-\nabla^2 (H_0\hat{r}) + \vec{\nabla}( \vec{\nabla}\cdot H_0\hat{r})] + m_H^2 H^{0i} = g_H \bar{\psi} \sigma^{0i} \psi .
\end{split}
\end{equation}
Note that using the same vector calculus identity as before, the Laplacian terms cancel out.
Therefore, this reduces the tensor meson interaction to that of a contact (or point) coupling interaction.
\begin{equation}
    \boxed{
    m^2_HH_0(r) = g_H\rho_t(r)
    }
\end{equation}
This equation is equally valid for our Lagrangian, however we seek a propagating meson in our theory to keep in line with the propagating mesons in the standard FSUGold class of RMF models.
In a point-coupling RMF interaction \cite{Zhang:2020azr,Liu:2023}, this meson interaction may be interesting to investigate further.
As a result, we shall use the configuration for the Case -1, where we adopt $a,b=+1$ and $c=-1$.


\section{Dirac Equation}
To derive the appropriate Dirac equation lets consider the full Lagrangian, where $\mathcal{L}_0$ is the standard RMF Lagrangian with scalar and vector fields whose nature can be isovector, isoscalar, or both (e.g. FSUGold, DINO, etc).
Taking our previously determined values for $\Lambda$, we write the Lagrangian again here as
\begin{equation}
    \mathcal{L} = \mathcal{L}_0 + \Big(\frac{1}{8} H_{\mu \nu \sigma} H^{\mu \nu \sigma} - \frac{1}{2} m_{H}^2 H_{\mu \nu} H^{\mu \nu} + g_H \bar{\psi} \sigma^{\mu \nu} H_{\mu \nu} \psi\Big)
\end{equation}
The field equation for the nucleon is obtained from the Euler-Lagrange equations,

\begin{equation}
    \frac{\partial \mathcal{L}_0}{\partial \bar{\psi}} + g_{H} \sigma^{\mu \nu} H_{\mu \nu} \psi = 0
\end{equation}

The previous definition for $H^{\mu \nu} = (g^{\mu i} g^{\nu 0} - g^{\mu 0} g^{\nu i})H_0(r) \hat{r}_i$ leads to the following Dirac-like equation,

\begin{equation}
    \frac{\partial \mathcal{L}_0}{\partial \bar{\psi}} - 2i g_{H} H_0(r) (\bm{\alpha} \cdot \hat{\mathbf{r}}) \psi = 0
\end{equation}

where we used $\sigma^{\mu \nu} H_{\mu \nu} = -2 \sigma^{0i} H_0(r) \hat{r}_i$ and $\sigma^{0i} = i \vec{\alpha}$.

\begin{equation}
    \frac{\partial \mathcal{L}_0}{\partial \bar{\psi}} - 2i g_{H} H_0(r) 
    \begin{pmatrix}
        0 & \sigma \cdot \hat{r} \\
        \sigma \cdot \hat{r} & 0
    \end{pmatrix} \psi = 0
\end{equation}

\begin{equation}
    \begin{pmatrix}
        E^*(r) - M^*(r) & i(\vec{\sigma} \cdot \vec{\nabla} ) \\
        -i(\vec{\sigma} \cdot \vec{\nabla}) & -E^*(r) - M^*(r) 
    \end{pmatrix}
    \psi
     - 2i g_{H} H_0(r) 
    \begin{pmatrix}
        0 & \sigma \cdot \hat{r} \\
        \sigma \cdot \hat{r} & 0
    \end{pmatrix} \psi = 0
\end{equation}
This reduces to two equations for the upper and lower components of the Dirac spinor:
\begin{align}
    \left( E^*(r) - M^*(r) \right) \frac{g_{n \kappa}(r)}{r} \mathcal{Y}_{\kappa m}(\mathbf{\hat{r}}) - (\vec{\sigma} \cdot \vec{\nabla}) \frac{f_{n \kappa}(r)}{r} \mathcal{Y}_{-\kappa m}(\mathbf{\hat{r}}) + 2g_H H_0(r) (\vec{\sigma} \cdot \hat{r}) \frac{f_{n \kappa}(r)}{r} \mathcal{Y}_{-\kappa m}(\mathbf{\hat{r}}) &= 0 \\
    \left( E^*(r) + M^*(r) \right) \frac{f_{n \kappa}(r)}{r} \mathcal{Y}_{-\kappa m}(\mathbf{\hat{r}}) + (\vec{\sigma} \cdot \vec{\nabla}) \frac{g_{n \kappa}(r)}{r} \mathcal{Y}_{\kappa m}(\mathbf{\hat{r}}) + 2g_H H_0(r) (\vec{\sigma} \cdot \hat{r}) \frac{g_{n \kappa}(r)}{r} \mathcal{Y}_{\kappa m}(\mathbf{\hat{r}}) &= 0
\end{align}

Using the recurrence and differential relations for spherical spinors \cite{Szmytkowski2006}, namely:
\begin{align}
    &(\vec{\sigma} \cdot \mathbf{\hat{r}}) \mathcal{Y}_{\kappa m}(\mathbf{\hat{r}}) = - \mathcal{Y}_{-\kappa m}(\mathbf{\hat{r}}) \\
    &(\vec{\sigma} \cdot \vec{\nabla}) F(r) \mathcal{Y}_{\kappa m}(\mathbf{\hat{r}}) = - \left( \frac{\partial}{\partial r} + \frac{\kappa + 1}{r} \right) F(r) \mathcal{Y}_{-\kappa m} (\mathbf{\hat{r}})
\end{align}
we get:
\begin{align}
    \left( \frac{d}{dr} - \frac{\kappa}{r} - 2g_H H_0(r) \right) f_{n \kappa}(r) + \left( E^*(r) - M^*(r) \right) g_{n \kappa}(r) = 0 \\
    \left( \frac{d}{dr} + \frac{\kappa}{r} + 2g_H H_0(r) \right) g_{n \kappa}(r) - \left( E^*(r) + M^*(r) \right) f_{n \kappa}(r) = 0 \\
\end{align}
For $E^*(r) = E - V(r)$ and $M^*(r) = M - S(r)$, we can write the coupled equations as in \cite{salinas:2024} with $T(r) = 2 g_H H_0(r)$
\begin{equation}
\boxed{
\begin{aligned}
    &\left( \frac{d}{dr} + \frac{\kappa}{r} + T(r) \right) g_{n \kappa}(r) - \big( E + M - V(r) - S(r) \big) f_{n \kappa}(r) = 0 \\
    &\left( \frac{d}{dr} - \frac{\kappa}{r} - T(r) \right) f_{n \kappa}(r) + \big( E - M - V(r) + S(r) \big) g_{n \kappa}(r) = 0
\end{aligned}
}
\end{equation}


\section{Contribution to Nuclear Binding Energy}
The contribution to the binding energy is found from the $00$-component of the stress-energy tensor.
Using \cite{salinas:2024} as reference, we can write the energy density as:
\begin{equation}
    \epsilon = \epsilon_0 - \left(\frac{1}{8} H_{\mu \nu \sigma} H^{\mu \nu \sigma} - \frac{1}{2} m_H^2 H_{\mu \nu} H^{\mu \nu}\right)
\end{equation}
where $\epsilon_0$ is the energy density from \cite{salinas:2024}. Using $H^{\mu \nu} = (g^{\mu i} g^{\nu 0} - g^{\mu 0} g^{\nu i})H_0(r) \hat{r}_i$ as before we get the following:
\begin{equation}
    \epsilon = \epsilon_0 - \left(\frac{1}{8} \left[ \partial_i H_{\nu \sigma} \partial^i H^{\nu \sigma} - 2\partial_i H_{j0} \partial^j H^{0i} \right] + m_H^2 H_0(r)^2\right)
\end{equation}
This reduces to,
\begin{align}
    \epsilon &= \epsilon_0 - \left(\frac{1}{8} \left[ -6 \partial_i H_0(r) \partial^i H_0(r) -2 \partial_i H_{j0} \partial^j H^{0i} \right] + m_H^2 H_0(r)^2 \right)\\
    &= \epsilon_0 - \left(\frac{1}{8} \left[6 (\vec{\nabla} H_0(r))^2 + 2 (\vec{\nabla} H_0(r) \cdot \hat{r}) (\vec{\nabla} H_0(r) \cdot \hat{r}) \right] + m_H^2 H_0(r)^2\right) .
\end{align}
Since $H_0(r)$ is only radially dependent, we may make use of the identity $(\vec{\nabla} H_0(r) \cdot \hat{r}) (\vec{\nabla} H_0(r) \cdot \hat{r}) = (\vec{\nabla} H_0(r))^2$ and we write:
\begin{equation}
    \epsilon = \epsilon_0 - \left((\vec{\nabla}H_0(r))^2 + m_H^2 H_0(r)^2\right)
\end{equation}

To get the total Energy, we integrate over the volume.
\begin{equation}
    E = E_0 - \int \left( (\vec{\nabla}H_0(r))^2 + m_H^2 H_0^2(r) \right) dV
\end{equation}
We can use the field equation in \cref{eq:proca-eq} to replace the mass term, $m_H^2 H_0^2(r) = \rho_t g_H H_0(r) - \frac{2}{r^2} H^2_0(r) + H_0(r) \nabla^2 H_0(r)$, resulting in the following integral:
\begin{align}
    E &= E_0 - \int \left( (\vec{\nabla}H_0(r))^2 + g_H \rho_t H_0(r) - \frac{2}{r^2} H_0^2(r) + H_0(r) \nabla^2 H_0(r)  \right) dV \\
    &= E_0 - \int \left( \vec{\nabla} \cdot (H_0(r) \vec{\nabla} H_0(r)) + g_H \rho_t H_0(r) - \frac{2}{r^2} H_0^2(r) \right) dV.
\end{align}
The first term in the integral vanishes from the divergence theorem, therefore the total contribution to the energy is:

\begin{align}
    \boxed{
   E = E_0 - 4 \pi \int \left( g_H \rho_t H_0(r) - \frac{2}{r^2} H_0^2(r) \right) r^2 dr
    }.
\end{align}

\newpage
\section{Green's Function}
Referring to the field equation in \cref{eq:proca-eq} for the tensor meson, we have:
\begin{equation}
\left(m_H^2 + \frac{2}{r^2} - \frac{2}{r} \frac{\partial}{\partial r} - \frac{\partial^2}{\partial r^2} \right)H_0(r) = g_H \rho_T(r).
\end{equation}
To obtain the Greens function we must find a function $D(r,r')$ such that
\begin{equation}
\left(m_H^2 + \frac{2}{r^2} - \frac{2}{r} \frac{\partial}{\partial r} - \frac{\partial^2}{\partial r^2} \right)D(r,r') = \delta(r-r').
\end{equation}
We can rearrange the differential equation into something more recognizable by examining the homogeneous case and multiplying through by $-r^2$.
\begin{equation}
r^2\frac{\partial^2 D(r,r')}{\partial r^2} + 2r \frac{\partial D(r,r')}{\partial r} - (m_H^2 r^2 + 2)D(r,r')  = 0
\end{equation}
Making the substitution $x = imr$ we can recast the differential equation into the general form of the spherical Bessel equation,
\begin{equation}
x^2\frac{\partial^2 D(x,x')}{\partial x^2} + 2x \frac{\partial D(x,x')}{\partial x} + (x^2 - 2)D(x,x')  = 0
\end{equation}
whose solution is a linear combination of spherical Bessel functions
\begin{equation}
    D(x,x') = A(x') j_1(x) + B(x') y_1(x).
\end{equation}

We now start to solve for $D(x,x')$ by using the inhomogeneous solution to obtain the left and right Green's function. 
We know that the green's function must be finite at the origin, and since $y_1(x)$ diverges as $x \rightarrow 0$, $B(x')=0$. 
The left Green's function is then simply,
\begin{equation}
    D_L(x,x') = A(x') j_1(x)
\end{equation}
Similarly for the right Green's function we have,
\begin{equation}
    D_R(x,x') = C(x') j_1(x) + E(x') y_1(x)
\end{equation}
Since we know the field must vanish as $r \rightarrow \infty$,
\begin{equation}
    \lim_{x \rightarrow i\infty} D_R(x,x') = 0.
\end{equation}
This then lets us solve for $C(x')$ in terms of $E(x')$:
\begin{equation}
\begin{split}
&C(x') j_1(x) + E(x') y_1(x) = 0 \\
&C(x') j_1(x) = - E(x') y_1(x) \\
&C(x') = - E(x') \frac{y_1(x)}{j_1(x)} \\
&C(x') = E(x') \frac{\frac{\cos x}{x^2} + \frac{\sin x}{x}}{\frac{\sin x}{x^2} - \frac{\cos x}{x}} \\
&C(x') = E(x') \frac{\cos x + x \sin x}{\sin x - x \cos x} .
\end{split}
\end{equation}
Since $x=imr$, we change variables again to $z=-ix$ before taking finally taking the limit as $z\longrightarrow\infty$. 
\begin{equation}
\begin{split}
&C(x') = E(x') \frac{\cos(iz) + iz \sin(iz)}{\sin(iz) - iz \cos (iz)} \\
&C(x') = E(x') \frac{\cosh z - z \sinh z}{i \sinh z - iz \cosh z} \\
&C(x') = -iE(x') \frac{\cosh z - z \sinh z}{\sinh z - z \cosh z} \\
&C(x') = -iE(x') \frac{\cosh z - z \sinh z}{\sinh z - z \cosh z} \\
&C(x') = -iE(x') \frac{e^z + e^{-z} - z e^z + z e^{-z}}{e^z - e^{-z} - z e^z - z e^{-z}}
\end{split}
\end{equation}
Taking the limit $z \rightarrow \infty$, we get,
\begin{equation}
    C(x') = -i E(x')
\end{equation}
The right Green's function is then,
\begin{equation}
\begin{split}
    D_R(x,x') &= -iE(x') j_1(x) + E(x') y_1(x) \\
    &= E(x')(y_1(x) - i j_1(x))
\end{split}
\end{equation}

Matching the left and right solutions at $x=x'$
\begin{equation}
\begin{split}
    &D_L(x',x') = D_R(x',x') \\
    &A(x') j_1(x') = E(x')(y_1(x') - i j_1(x')) \\
    &A(x') = E(x') \frac{y_1(x') - i j_1(x')}{j_1(x')}
\end{split}
\end{equation}
gives us the following left and right Green's functions with one unknown,
\begin{equation}
\begin{split}
    &D_L(x,x') = E(x') \frac{y_1(x') - i j_1(x')}{j_1(x')} j_1(x) \\
    &D_R(x,x') = E(x')(y_1(x) - i j_1(x))
\end{split}
\end{equation}
The function $E(x')$ is obtained by integrating over the differential equation about $r'\pm\epsilon$:
\[
\int_{r'-\epsilon}^{r'+\epsilon} \left(m_H^2 + \frac{2}{r^2} - \frac{2}{r} \frac{\partial}{\partial r} - \frac{\partial^2}{\partial r^2} \right)D(r,r') dr = \int_{r'-\epsilon}^{r'+\epsilon}  \delta(r-r') dr \\
\]
\[
\int_{r'-\epsilon}^{r'+\epsilon} m_H^2 D(r,r') dr + \int_{r'-\epsilon}^{r'+\epsilon} \frac{2}{r^2} D(r,r') dr - \int_{r'-\epsilon}^{r'+\epsilon} \frac{2}{r} \frac{\partial D(r,r')}{\partial r} dr - \int_{r'-\epsilon}^{r'+\epsilon} \frac{\partial^2 D(r,r')}{\partial r^2} dr = 1
\]
The first two integrals vanish since $D(r,r')$ is continuous and finite at $r=r'$, leaving us with:
\[
- \int_{r'-\epsilon}^{r'+\epsilon} \frac{2}{r} \frac{\partial D(r,r')}{\partial r} dr - \int_{r'-\epsilon}^{r'+\epsilon} \frac{\partial^2 D(r,r')}{\partial r^2} dr = 1
\]

\[
\int_{r'-\epsilon}^{r'+\epsilon} \frac{2}{r} dD(r,r') + \int_{r'-\epsilon}^{r'+\epsilon} d \frac{d D(r,r')}{dr} = -1
\]

\[
\int_{r'-\epsilon}^{r'+\epsilon} \frac{2}{r} dD(r,r') + \int_{r'-\epsilon}^{r'+\epsilon} d \frac{d D(r,r')}{dr} = -1
\]

\[
\frac{2}{r} D(r,r') \bigg{\vert}_{r'-\epsilon}^{r'+\epsilon} + \frac{d D(r,r')}{dr} \bigg{\vert}_{r'-\epsilon}^{r'+\epsilon} = -1
\]

\[
\frac{2}{r'+\epsilon} D_R(r'+\epsilon,r') - \frac{2}{r'-\epsilon} D_L(r'-\epsilon,r') + \frac{d D(r,r')}{dr} \bigg{\vert}_{r'-\epsilon}^{r'+\epsilon} = -1
\]
As $\epsilon \rightarrow 0$, the LHS vanishes since $D_R(r',r') = D_L(r',r')$. Thus we are left with,
\[
\frac{d D(r,r')}{dr} \bigg{\vert}_{r'-\epsilon}^{r'+\epsilon} = -1
\]

\[
\frac{d D_R(r,r')}{dr} \bigg{\vert}_{r'+\epsilon} - \frac{d D_L(r,r')}{dr} \bigg{\vert}_{r'-\epsilon} = -1
\]
Recasting $x=imr$, we have
\begin{equation}
\frac{d D_R(x,x')}{dx} \bigg{\vert}_{x'+\epsilon} - \frac{d D_L(x,x')}{dx} \bigg{\vert}_{x'-\epsilon} = \frac{i}{m}.
\end{equation}
Using the expressions for $D_L(x,x')$ and $D_R(x,x')$,

\begin{equation}
\begin{split}
    &D_L(x,x') = E(x') \frac{y_1(x') - i j_1(x')}{j_1(x')} j_1(x) \\
    &D_R(x,x') = E(x')(y_1(x) - i j_1(x))
\end{split}
\end{equation}
we get the following equation for $E(x')$ where the prime on the Bessel function denotes the derivative
\begin{equation}
    \begin{split}
        E(x') &= \frac{i}{m} \frac{j_1(x')}{y_1'(x') j_1(x') - y_1(x') j_1'(x')}\\
        &= \frac{j_1(x')}{y_0(x') j_1(x') - y_1(x') j_0(x')} \\
        &= \frac{i}{m}(\sin x' - x' \cos x') \\
        &= \frac{i}{m} x'^2 j_1(x')
\end{split}
\end{equation}
Thus, the expressions for the left and right Green's functions are
\begin{align}
    D_L(x,x') &= \frac{i}{m}x'^2 \left[ y_1(x') - i j_1(x') \right] j_1(x) \\
    D_R(x,x') &= \frac{i}{m}x'^2 \left[ y_1(x) - i j_1(x) \right] j_1(x') .
\end{align}
Substituting back in $x=irm$ and using the spherical Hankel functions for the bracketed terms, gives this expression for the left and right greens function.
\begin{align}
    D_L(r,r') &= -\frac{i}{m} e^{-mr'} (1+mr') j_1(imr) \\
    D_R(r,r') &= -\frac{i}{m} \left( \frac{r'}{r} \right)^2 e^{-mr} (1+mr) j_1(imr') \\
\end{align}
or in terms of transcendental functions,
\begin{align}
    D_L(r,r') &= \frac{e^{-mr'}}{m^2r} (1+mr') \left[ \cosh(mr) - \frac{\sinh(mr)}{mr} \right] \\
    D_R(r,r') &= \frac{e^{-mr}}{m^2r'} (1+mr) \left( \frac{r'}{r} \right)^2 \left[ \cosh(mr') - \frac{\sinh(mr')}{mr'} \right].
\end{align}
In compact notation, by defining the Greens function over the full volume, the final Green's function is
\begin{equation}
    \boxed{
    D(r,r';m)=\frac{e^{-mr_>}}{4 \pi m^2r_>^2 r_<}(1 + mr_>) \left[ \cosh(mr_<) - \frac{\sinh(mr_<)}{mr_<} \right].
    }
\end{equation}

\bibliography{bib}
\end{document}